\documentclass[12pt]{article}
\usepackage{epsfig}

\def\lll{\langle}
\def\T{\mbox{Tr}\>}
\def\P{\mbox{P}\exp}
\def\al{\alpha}
\def\nn{\nonumber}
\def\rrr{\rangle}

\def\fun#1#2{\lower3.6pt\vbox{\baselineskip0pt\lineskip.9pt
\ialign{$\mathsurround=0pt#1\hfil
##\hfil$\crcr#2\crcr\sim\crcr}}}
\newcommand{\be}{\begin{equation}}
\newcommand{\ee}{\end{equation}}
\newcommand{\beqa}{\begin{eqnarray}}
\newcommand{\enqa}{\end{eqnarray}}

\title{Field correlators in QCD.\\
Theory and applications.}
\author{
A.Di Giacomo\thanks{digiaco@mailbox.difi.unipi.it}\\
{\it\small Pisa University and INFN Sezione di Pisa, Italy},\\
H.G.Dosch\thanks{H.G.Dosch@thphys.uni-heidelberg.de}\\
 {\it\small Institut f\"ur Theoretische
Physik der Universit\"at Heidelberg, Germany}, \\   
V.I.Shevchenko\thanks{corresponding author, e-mail: V.Shevchenko@phys.uu.nl} \\
{\it\small Institute for Theoretical Physics, Utrecht University, Netherlands} \\
{\small and} \\
{\it\small Institute of
Theoretical
and Experimental
Physics, Moscow, Russia}, \\ 
Yu.A.Simonov\thanks{simonov@heron.itep.ru}\\
{\it\small Institute of
Theoretical
and Experimental
Physics, Moscow, Russia}
}
\begin{document}
\maketitle

\newpage 

\vspace{4cm}

\begin{abstract}
This review is aimed to demonstrate 
the basics and use of formalism of gauge-invariant nonlocal 
correlators  in nonabelian gauge theories. 
Many phenomenologically interesting 
nonperturbative aspects of 
gluodynamics and QCD can be described in terms of  
correlators of the nonabelian field strength tensors.
It is explained how the properties of correlator ensemble 
encode the structure of QCD vacuum and determine 
different nonperturbative observables. It is argued that in gluodynamics 
and QCD the dominant role is played by the lowest nontrivial two-point correlator
(Gaussian dominance). Lattice measurements of field correlators
are discussed. Important for the formalism theoretical tools, 
such as nonabelian Stokes
theorem, background perturbation theory, cluster expansion, as
well as phenomenological applications to the heavy quarkonium 
dynamics and QCD phase transition are reviewed.
\end{abstract}

\bigskip
\bigskip


\bigskip
\bigskip


\newpage

\tableofcontents

\newpage

\section {Introduction and motivation}

\setcounter{equation}{0}
\def\theequation{1.\arabic{equation}}

 This review is devoted to the study of gauge theories, specifically
 gluodynamics and ${\mbox{QCD}}_4$. There are now many good books on QCD
\cite{books1,books2,books3,books4}, which mainly concentrate on perturbation theory. The
 nonperturbative (NP) aspects of QCD (and gauge theories in general)
 have a somewhat peculiar status. On the one hand our very  existence
 is based on one of the most important NP phenomenon --
 confinement.
 On the other hand  perturbative QCD
 disregards confinement and treats quarks
 and gluons as weakly interacting.
 Perturbative QCD  is
 a vast field of investigations which makes most of the contents
 of the QCD textbooks leaving for the NP part of QCD only few
 short qualitative chapters at the end.

  The main unsolved  problem of QCD is the quantitative treatment of
 NP phenomena.
We still do not have selfconsistent formalism where nonperturbative
 quantities,  like masses and condensates, are computed
 systematically from first principles, i.e. from the QCD
 Lagrangian. Nowadays the only reliable way of getting NP results 
in QCD is
numerical simulation of the theory on the lattice. This area of
research is of great importance, bringing  
many insights into the subject \cite{rev21,rev22}.

 Another source of the NP intuition in QCD
is the analysis of quantum
field theoretical models simplified in one respect or another.
These are a huge zoo of low-dimensional
models (notably ${\mbox{QCD}}_2$),
abelian systems, supersymmetric theories and many others.
The studies of the corresponding exact solutions 
provide substantial information 
relevant for physically more interesting cases.

We have  no experimental chance to study the properties of the
world without light quarks. However, lattice calculations and
theoretical analysis in the limits where quark degrees of freedom
are decoupled show, that the property of confinement (i.e. absence
of colored particles as asymptotic states 
\cite{RPP94}) holds in the theory without quarks (pure gluodynamics) as
well as in real QCD. It suggests to study the physics of
confinement starting from quenched approximation, as 
is done, for example, in $1/N_c$ -- expansion. The asymptotic freedom
\cite{gvp1,gvp2}  and dimensional transmutation phenomena provide
the NP mass scale, $\Lambda_{QCD}$ and force the theory to
leave its perturbative domain at large distances. All other
dimensionful quantities (such as string tension for static
external sources, gluon condensate, glueball and hybrid masses and
widths) in the quenched approximation are proportional to
corresponding power of 
$\Lambda_{QCD}$ with some dimensionless coefficients
that in principle can be computed in NP QCD.

The chiral effects, which are mostly responsible for
physics of lightest hadrons, are originated from
the NP gluonic sector of the theory as well since
there is no spontaneous chiral symmetry breaking in
perturbative QCD.
The confinement -- chiral
symmetry breaking correspondence, however, is
very far from trivial. We do not touch this set of questions
in this review, planning to present an extended separate discussion
of this subject elsewhere. An interested reader can find
a lot about the light quark physics in the confining
vacuum in the gauge-invariant field correlators
approach in \cite{chir}.

An important question is related with possible phases of the
theory and, in particular, with the NP effects in different
phases. Internal parameters of the theory (such as number of
colours and flavours, light quark masses) and external conditions
(temperature, chemical potential, number of space-time dimensions)
both play important role here. They determine the symmetry of the
ground state at a given point of the phase diagram and
corresponding spectrum of light excitations.
Each point of the phase diagram can be characterized
by the values of condensates, describing the symmetry breaking
pattern. The simplest condensates are given by the NP gluon and
quark condensates:
\be
\lll \frac{\alpha_s}{\pi} F_{\mu\nu} F_{\mu\nu} \rrr \;\;\; ;
\;\;\;\; \lll \bar\psi \psi \rrr \label{eq1} \ee The average in
(\ref{eq1}) is defined by (\ref{2.5}) below and all perturbative
contributions are assumed to be subtracted from (\ref{eq1}). The
NP--generated nonzero value of the gluon condensate \cite{npgl}
corresponds to
the breaking of scale invariance at the quantum 
level (notice, that finite
dimensionful
parameter cannot be generated perturbatively). Nonzero value of
the quark condensate indicates chiral symmetry breaking
\cite{books1,books4}. One can also consider different higher order
operators as well as nonlocal averages. In the high density -- low
temperature limit the nonzero diquark condensate $\lll \psi\psi
\rrr$ is predicted, leading to the phenomenon of colour
superconductivity, while chiral symmetry is restored
\cite{colsup1,colsup2,colsup3}. The actual values of the condensates also depend on
the temperature and another parameters of the theory.

 As it has already been mentioned,
physics of light quarks will not be discussed
in the present review. Our main interest
is concentrated on a systematic method of
computing NP effects in the quenched approximation
from some fundamental NP input, one and the same for
all processes.
We have chosen for this NP input a
 set of field strength correlators (FC), namely, \be
\Delta_{\mu_1\nu_1,...,\mu_n\nu_n} = \T \lll F_{\mu_1\nu_1} (x_1) \Phi
(x_1x_2)
 F_{\mu_2\nu_2}(x_2)...  F_{\mu_n\nu_n} (x_n) \Phi (x_nx_1) \rrr \label{eq3}
 \ee
where
$F_{\mu\nu}$'s are the field strength tensors and $ \Phi(x,y)$ are
 the phase factors, introduced for the sake of gauge invariance
 and discussed in details below.

The systematic use of gauge-invariant variables was adopted for the first time by
S.Mandelstam in his celebrated paper {\cite{mand}}. It was shown, that
usual Feynman
rules of perturbation theory can be recovered from the
Dyson-Schwinger equations for the averages (\ref{eq3}).
The key idea proposed in \cite{ds11,ds12,ds13} is
to use the
gauge-invariant quantities (\ref{eq3})
as a dynamical input also in the nonperturbative domain
and to express
gauge-invariant observables through (\ref{eq3}) via the cluster expansion.
Research activity during the last 10 years has shown that several
QCD processes of phenomenological interest
can be calculated using the set (\ref{eq3}) as an input.
 More than that, a systematic cluster expansion 
can be performed, and the first term, corresponding
to the FC with $n=2$ (sometimes called
 later the Gaussian
 or bilocal correlator) already gives a good 
qualitative description of most NP
 phenomena, while higher cumulants can be considered 
as corrections (see Section 2).
 There are evidences that these corrections in the known cases are not
 large and contribute around a few percents of the total effect. Thus one
 obtains a theory with a simple but fundamental input  --  the
 Gaussian correlator --  and the corresponding formalism is called
 the Gaussian dominance approximation or sometimes the Gaussian stochastic
 model of QCD vacuum.

The method discussed in the review can be called "fundamental
phenomenology" since it uses correlators (actually the
lowest Gaussian one) (\ref{eq3}) as the only
dynamical input. The latter is usually given by lattice
measurements. It is worth mentioning that actual 
profile of the Gaussian correlator plays no essential role for the
calculation of hadron properties. The necessary NP information
enters via string tension $\sigma$, which is an integral characteristic
of Gaussian correlator and thus one has a unique opportunity
to define the hadron spectrum in terms of one parameter.
Note also in this respect, that we do not
try to reconcile the present approach  with any of the
particular NP QCD vacuum microscopical models, in particular,
instanton liquid model in this review (see e.g. \cite{revconf}).
As it will be clear from what follows, since the quantities
(\ref{eq3}) are eventually the main input objects,
one might not in principle be interested what kind of field
configurations play the dominant role for
the particular behaviour of correlators.
On the other hand, most of the dynamical statements of
a particularly chosen model about the NP content of the QCD vacuum
can typically be reformulated as a gauge--invariant
statements about properties of the correlators (\ref{eq3}),
especially if the model provides some recipee to
calculate them nonperturbatively (see also
discussion in \cite{dima2}).

Lattice measurement is a very important source  of information on FC.
First calculation of the Gaussian FC in
\cite{lat11,lat12}  was followed
by more detailed investigation in \cite{lat2,lat31,lat32} where also nonzero
temperature was explored. Both Gaussian and quartic FC have been
measured while testing the field distribution inside the QCD string
\cite{lat41,lat42,lat43}  and this analysis together with independent measurement
done in \cite{bbv} supports the Gaussian dominance picture.

 One should say a few words about the symmetries of QCD
vacuum, which make it natural to use FC given by (\ref{eq3}). 
Since the fields $A_{\mu}$ as well as ${\vec{E}}_i$
and ${\vec{B}_i}$ are vector--like and Lorentz invariance and color
neutrality of the vacuum lead to inevitable conclusion that
there is no any preferred direction either in the coordinate or
in the color space in the physical vacuum of QCD, one immediately 
gets 
\be
\lll F_{\mu\nu}(x) \rrr \equiv 0
\ee
This rules out models with constant vector fields.  
There is an important difference with the scalar field 
condensation in Higgs-like theories.
The latter is usually characterized by some constant
parameter, e.g. value of the scalar field in unitary gauge.
This is impossible for the vector field condensation
since the construction of true vacuum state implies an
averaging procedure over different field configurations 
in order to get Lorentz and gauge invariant ground state.
 These averaging algorithms, however, may be of different types:
for example, in models based on ensemble of classical
field configurations like instanton gas model one performs 
it by summing over all positions and colour 
orientations of individual instantons, while in the original 
quantum case the summation over 
properly weighted gauge field configurations is "built in" from the 
beginning in the functional integral approach.

It is known already for more than 20 years \cite{npgl} that the QCD
vacuum is filled with
strong NP chromoelectric (CE) and chromomagnetic (CM) fields, which due
to the  scale anomaly cause the NP shift of the energy density of the
 vacuum \cite{books1}
 \be
 \varepsilon =\frac{\beta(\alpha_s)}{16\alpha_s}
 \lll F_{\mu\nu}(0) F_{\mu\nu}(0) \rrr
\label{eq5}
\ee
 Because of the asymptotic freedom $\beta(\alpha_s)$ is negative (at
 least for small $\alpha_s)$  thus
 making the NP shift $\varepsilon$
 energetically favorable.

Now a crucial question arises, what are the typical scales of the
fields, contributing to the r.h.s. of (\ref{eq5})?  The original proposal
made in \cite{npgl} considered these fields as slowly varying in
space-time and hence, in the operator product expansion (OPE) formalism
the condensates enter as constant coefficients in front of the terms with
powers of momentum while higher twist
contributions are suppressed as
subleading corrections.

However, in the course of lattice
simulations it was found \cite{lat11,lat12}, that typical correlation length $T_g$,
characterizing the spatial dependence of 
gauge-invariant bilocal correlator
(\ref{eq3}) is very small, of the order of 0.2-0.3 fm (see
discussion below).
It seems therefore very natural to expect the onset of
conventional OPE at such large $Q^2$ that 
$Q^2 >> T_g^{-2}$, while at $Q^2 \simeq
T_g^{-2}$ the system crucially feels the 
nontrivial space-time
distribution of the fields
and usual OPE can break down.

One can  easily see that the method of field
 correlators (MFC) described in the review is a natural development
 of the QCD sum--rule method, where the NP inputs are
 condensates, i.e. vacuum averages of operator products taken at one
 point. The sum--rule method has natural limits and cannot be applied to
 long distance processes, since information stored in local averages
 is not enough to describe dynamics in the region where  stochastic
 nature of the vacuum plays the most important role.
Thus higher excited hadron states, string--like spectrum and all
 properties connected to confinement are outside of realm of the  QCD
 sum--rule method.

Another essential ingredient of the method is the idea of Gaussian
dominance.
As we shall discuss in Section 3, the Gaussian approximation is a
 very specific model, where all higher correlators (\ref{eq3}) are
 expressible through the product of lowest Gaussian ones.
 Then all  correlators with odd $n$ are assumed to vanish. A
 less restrictive approximation is the one where the correlators
 with $n=2,3$ are nonzero and the higher FC factorize into
 product of lowest ones. This  approximation is sometimes called the
 extended Gaussian model.
It is worth noting that the term "Gaussian" should not be
misunderstood. The theory under consideration is not a free
theory. The $n$ - point Green function is assumed to be
factorized as a product of bilocal ones, which
however are not propagators of free fields. The
Gaussian dominance has essentially dynamical origin, which is not
clear yet, but is strongly supported by lattice calculations and
hadron phenomenology. As was mentioned above the Gaussian model is
quite successful in decribing NP dynamics in all situations studied
heretofore. Corrections due to higher correlators in
the cases studied were not large (about a few percent for the
static potential).

 Contrary to the QCD sum--rules MFC is applicable both at small and
 large distances. It includes perturbative QCD at small  distances with
 asymptotic  freedom, but also describes the extrapolation to large
 distances.

 The MFC has been applied to describe the  hadronic
 spectrum for light quarks \cite{lq}, heavy quarkonia \cite{hq1,hq2,hq3,hq4,hq5},
 heavy-light mesons \cite{hl},
glueballs and hybrids \cite{glhb1,glhb2},
formfactors and structure functions in
 the proper light--cone dynamics \cite{lc1,lc2},
string formation \cite{stfrm},
 diffractive photoproduction \cite{dif},
scattering \cite{sc11,sc12,sc13,sc21,sc22,sc23} and also
 outside of QCD to calculate the spectrum of electroweak theory
 \cite{ew}.

 There exist several reviews which are partly devoted to specific
 applications of MFC, namely, \cite{rev1} treats mostly the hadron
 spectrum, scattering is discussed in \cite{revsc1,revsc2,revsc3}, the
 problem of confinement is reviewed in \cite{revconf}, and  QCD at nonzero
 temperature and deconfinement in \cite{revtc}, brief overview
of MFC is presented in \cite{shladm}.

Another example of applications is the calculation of FC directly
in the framework of a given field--theory, which is possible in
the Abelian Higgs model and compact electrodynamics (see
detailed review \cite{dima2} and references therein).

 We do not touch most of these applications in the present review,
 referring the reader to the cited literature. Instead, we
 concentrate
 here on the general  and more fundamental questions, including
 definition of FC in the continuum and on the lattice, the
 renormalization, connection to OPE, low--energy theorems etc.
We discuss the
 basic properties of NP QCD as a testing ground for the method,  and
 show  how they are described in the method of field
correlators already in its simplest,
Gaussian approximation.

The plan of the review is the following. Section 2
is devoted to the basic formalism of the method: definitions of field
correlators, their elementary properties, factorization, nonabelian Stokes
theorem. The physics of confinement in the language of field correlators
is discussed in details in Section 3: cluster expansion,
string formation and profile, perturbative properties,
low energy theorems, confinement
of the charges in higher representations of the gauge group and
some other questions are reviewed. Section 4 treats the physics of
heavy quarkonia by the method of field correlators. Finally, the
deconfinement
transition and aspects of NP physics at nonzero temperature are
discussed in Section 5.

\newpage

\setcounter{section}{1}
\section {Basic formalism}

\setcounter{equation}{0} \def\theequation{2.\arabic{equation}}

In this Section we discuss some basic definitions
and general formalism of nonlocal gauge-invariant
field correlators as applied to $SU(N)$ Yang-Mills
theory.

\bigskip

\subsection{Definition of field correlators}

\bigskip

 For the most part of the review we are working in the Euclidean
 formalism and consider the Euclidean vacuum picture of QCD fields.
 From the QCD sum rules
and heavy quarkonia spectrum analysis
 we know that the gluon
vacuum is dense,
\be 
G_2 \equiv \frac{\alpha_s}{\pi} \lll F^a_{\mu\nu}F^a_{\mu\nu} \rrr
\sim  0.012\> \mbox{GeV}^4.
\label{2.1}
 \ee

A dynamical characteristic of such stochastic vacuum is given by a
set of gauge-invariant v.e.v., which in the
nonabelian case have the form
\be
\frac{1}{N_c} \> \lll \T G_{\mu_1\nu_1}(x_1, x_0)
G_{\mu_2\nu_2}(x_2, x_0)
...
G_{\mu_n\nu_n}(x_n, x_0) \rrr =
\Delta_{1,2...n}
\label{2.4}
\ee
where
\be
G_{\mu_k\nu_k}(x_k, x_0)=
\Phi(x_0, x_k) F_{\mu_k\nu_k}(x_k)\Phi(x_k, x_0)
\label{eq234}
\ee
and phase factors $\Phi$ are defined as follows
\be
\Phi(x,y)= \P i \int^x_y A_{\mu} dz_\mu \label{2.3} \ee with the
integration going along some curve, connecting the initial and the
final points. The angular brackets are defined here as vacuum
expectation value with the usual QCD weight, the
exact form of which is given by
\be
\lll {\cal Q}\rrr =\int d\mu(A)e^{-S(A)}{\cal Q}(A)
\label{2.5}
\ee
Here $S(A)$ is the standard Yang-Mills action
$$
S(A) = \frac{1}{2g^2} \int d^4 x \>\T F_{\mu\nu}^2
$$
and the measure $ d\mu(A)$ includes gauge-fixing and
the corresponding Faddeev-Popov terms.
The fermion determinant is taken to be unity in the
quenched case considered here.

The most attractive feature of the nonlocal averages
(\ref{2.4}) is their gauge-invariance, as compared with the case of
usual gauge field Green's functions
$\lll A(x)A(y)..A(z)\rrr $.
There is a serious price to pay, however --- functions $
\Delta_{1,2...n}$ depend on the form of the contours, entering
(\ref{2.3}).
This dependence is to be cancelled in physical quantites,
and we will discuss below how such cancellation occurs.
It will be argued, in particular, that in Gaussian approximation
contour dependence is a subleading effect.

The form of basic quantities (\ref{2.4}) suggests
to use the special class of gauge conditions, the so called
generalized contour gauges \cite{cg,cg1}, to be discussed below.
Of particular importance is the case of straight-line contours
in (\ref{2.3}), which corresponds to Fock-Schwinger (F-S)
gauge condition \cite{fockscw1,fockscw2}.
In this case any $n$--point correlator can be
constructed using the invariant tensors
$\varepsilon_{\mu\nu\sigma\rho},  \delta_{\mu\nu}$,
vectors
{$(z_i-z_0)_{\mu}$} and several scalar functions,
depending on relative
coordinates $(z_i-z_0)^2, (z_i-z_j)^2$, where $z_0$ is the base point
of the F-S gauge, which can be placed at the origin.
The properties of these functions play a crucial
role in what follows (see Section 3).

For the simplest nontrivial 2--point correlator with the points
$z_1$ and $z_2$ connected by the straight line phase factors (i.e
the point $z_0$ is chosen on the line, connecting $z_1$ and $z_2$)
one parametrizes bilocal correlator conventionally as \cite{ds11,ds12,ds13}:
$$ \Delta^{(2)}_{\mu_1\nu_1,\mu_2\nu_2}= \frac{1}{N_c} \T \lll
(F_{\mu_1\nu_1}(z_1)\Phi(z_1,z_2)
F_{\mu_2\nu_2}(z_2)\Phi(z_2,z_1))\rrr =
  $$
 $$
= \; \frac12\> \left(\frac{\partial}{\partial z_{\mu_1}}
(z_{\mu_2} \delta_{\nu_1 \nu_2} - z_{\nu_2}
\delta_{\nu_1 \mu_2}) +
  \frac{\partial}{\partial z_{\nu_1}}
(z_{\nu_2} \delta_{\mu_1 \mu_2} - z_{\mu_2} \delta_{\mu_1
\nu_2})\right)\> D_1(z_1-z_2) +
  $$
\be
+ (\delta_{\mu_1\mu_2} \delta_{\nu_1\nu_2} - \delta_{\mu_1\nu_2}
\delta_{\mu_2\nu_1}) \> D(z_1 - z_2) \label{bil} \ee 
This
representation does not contain $CP$-noninvariant part,
proportional to ${\varepsilon}_{\mu_1\nu_1\mu_2\nu_2}$, which
should be included in theories with nonzero $\theta$-term.

Another useful representation comes if one considers correlators
of CE and CM fields separately : $$ \frac{1}{N_c}\>\T
\lll (E_{i}(x)\Phi(x,y) E_{j}(y)\Phi(y,x))\rrr = $$ $$ =
{\delta}_{ij} \left(D(z) + D_1(z) + [z_4]^2 \>\frac{dD_1(z)}{dz^2}
\right) + z_i z_j \frac{dD_1(z)}{dz^2} $$ $$ \frac{1}{N_c} \T \lll
(B_{i}(x)\Phi(x,y) B_{j}(y)\Phi(y,x))\rrr = $$
\be
= {\delta}_{ij} \left(D(z) + D_1(z) + [\vec z\,]^2\>
\frac{dD_1(z)}{dz^2} \right) - z_i z_j \frac{dD_1(z)}{dz^2}
\label{bil1} \ee where $z=x-y$. The following notation is useful
\cite{lat2}:
\be
D_{||}(z) = D(z) + D_1(z) + z^2
\frac{dD_1(z)}{dz^2} \;\;\;  ; \;\;\;
D_{\bot}(z) = D(z) + D_1(z)
\label{rel1}
\ee
The 2-point correlator of the dual field strength $\tilde F_{\mu\nu}
=\frac12 {\varepsilon}_{\mu\nu\alpha\beta} F_{\alpha\beta}$
is constructed from dual functions $\tilde D, {\tilde D}_1$ in complete
analogy with (\ref{bil}), the dual functions are related with $D$ and
$D_1$ according to
\be
{\tilde D}(z) =  D(z) + 2 D_1(z) + z^2\>
\frac{dD_1(z)}{dz^2} \;\;\; ;\;\;  {\tilde D}_1(z) = - D_1(z)
\label{rel2}
\ee
It is immediately seen, that on the (anti)-selfdual configurations
$D_1 \equiv 0$.

Let us discuss some basic properties of the functions introduced above.
First, consider the case of QED \cite{vza}.
The gauge field propagator in the Feynman gauge takes the form
\be
\lll eA_{\mu} (x) eA_{\nu} (y)\rrr =
{\delta}_{\mu\nu}\frac{1}{4\pi^2} \frac{e^2(z)}{z^2}
\label{es1} \ee where $z=x-y$ and the square of the invariant
charge $e^2(z)$ (in coordinate space) is introduced.
Differentiating both sides of (\ref{es1}) with respect to $z$ one
finds for $\lll F_{\mu\nu} F_{\rho\sigma} \rrr$ in QED:
\be
D(z) \equiv 0 \;\; ; \;\;\; D_1(z)
=-\frac{1}{\pi^2}\>\frac{d}{dz^2}\>\left(\frac{e^2}{z^2} \right)
\label{es2} \ee and
\be
\lll e^2 F_{\mu\nu}(x) F_{\mu\nu}(y) \rrr =
\frac{6}{{\pi}^2}\>
\frac{1}{z^4}\> \left( 1 -
\frac{d\beta(e^2)}{d e^2} \right)\cdot \beta(e^2)
\label{es3}
\ee
where
 $$ z^2 \frac{de^2}{dz^2} = \beta(e^2) $$

The formulas (\ref{es2})--(\ref{es3}) are obtained
in abelian theory. Notice that $D(z) \equiv 0$ to all orders
in perturbation theory.
The case of Yang-Mills theory is different from that mostly
in two respects: there are perturbative contributions to
the function $D(z)$ and, which is more important, both
functions $D$ and $D_1$
acquire NP parts, to be denoted $D^{pert}(z)$ and 
$D^{np}(z), D_1^{np}(z)$, respectively 
Pure perturbative contributions to $D$ and $D_1$ are
discussed in the Section 3, while most of the rest of the review
is devoted to the NP parts, and we usually omit the
subscripts $pert$ and $np$.   

The perturbation theory dictates 
the singular behaviour of $D^{pert} , D_1^{pert}$ at the origin:
$D^{pert}(z)\sim z^{-4}$, if $z^2\to 0$. The
NP part is normalized to the gluon
condensate (\ref{2.1}) according to:
\be
24 N_c \cdot(D^{np}(0) + D_{1}^{{np}}(0)) = \lll
F^a_{\mu\nu}F^a_{\mu\nu} \rrr \equiv 4 \pi^2 \> G_2
\label{eq34} \ee 
This information is to be used in fitting the lattice data 
on correlators, which we are now in the position to discuss.
The first measurements were made in quenched $SU(2)$ gauge theory
\cite{lat11,lat12}. The motivation was to correct the predictions of 
\cite{6.1,6.2}
of the shift produced by the gluon condensate on the levels of heavy
$\bar{q} q$ bound states: the introduction of a finite correlation length 
strongly reduces the effect on higher levels. 
The NP part exhibited an exponential dependence on the physical distance 
with a slope $T_g \approx 0.16 $ Fm. The condensate was also
determined with the result $G_2 = 0.012 {\mbox{GeV}}^4$. Of course the
energy scale is conventional, the string tension for $SU(2)$ being
put equal to the physical string tension $\sigma \approx \frac{1}{2\pi} 
\mbox{GeV}^2$ . What is actually measured are loops made of two 
plaquettes connected by the phase factors along straight lines,
with different orientation of the plaquettes. 
A progress was made a few years later \cite{lat2} 
when the correlators were measured for $SU(3)$ gauge theory 
in the quenched approximation, i.e. with no dynamical quarks. Besides the 
increase in the computer power available the progress was made possible 
by using cooling technique to study large distance correlations \cite{cool}.
The configurations produced numerically by use of the correct action 
are filtered by this procedure eliminating short range fluctuations,
but leaving at the same time intact long range correlations. The principle
is the following: cooling is done by a local procedure which freezes
singular links. It can be shown that the maximum distance 
$L$ which is affected after $n$ steps of cooling depends on $n$ in 
a diffusion-like form : $ L^2 = D\cdot n$. In measuring a correlation at distance 
$d$ a plateau is eventually observed after a number of cooling steps,
on which the value of the correlator is read, corrected for short distance
fluctuations. Empirically the procedure requires a minimum distance 
of 3-4 lattice spacings to work. This makes hard to measure correlators
at short distances. On the one hand the shortest distance observed 
(say 0.1 Fm) should correspond to 3-4 lattice spacings at least.
On the other hand to avoid dominance of surface effects the lattice
must be large enough to cover $\sim 1.5$ Fm in distance.
In \cite{lat31,lat32} the correlators was measured for quenched 
$SU(3)$ from 0.1 Fm to 1 Fm in distance.

At large distances an exponential fit correctly
describes the data. At short distances a divergent behavior
$\sim z^{-4}$ is observed as predicted by perturbation theory.
There is no theoretically rigorous way to separate these two contributions
on the lattice. Different parametrizations were used in \cite{lat2,lat32}
to fit the data and they proved to be equally good.
One possibility is   
$$ \frac{1}{a^4} \>
{D}^{lat}(z^2) = \frac{b}{|z|^4}\exp(-|z|/\lambda_a) + A_0
\exp(-|z|/T_g) $$
\be
\frac{1}{a^4} \> {D}^{lat}_{1}(z^2) =
\frac{b_1}{|z|^4}\exp(-|z|/\lambda_a) + A_1 \exp(-|z|/T_g)
\label{adg2} \ee
The constants $A_0 , A_1$ are related to the gluon 
condensate $G_2$ as follows (see (\ref{eq34})):
\be
A_0+ A_1 = \frac{\pi^2}{18}\> G_2
\label{conn}
\ee
A fit to the lattice data gives for quenched $SU(3)$
(see also Fig.1)
\be
T_g \approx 0.22 \mbox{Fm} \;\;\;\;\; ; \;\;\;\;\; 
G_2 = (0.14 \pm 0.02) {\mbox{GeV}}^{4} ,
\label{data1}
\ee
i.e. a value large by an order of magnitude 
than the phenomenological value of \cite{npgl}.
Moreover it was found $\left| A_1 / A_0 \right| \approx 0.1$.

In
\cite{lat31} the measurements were done 
in
full QCD with 4 staggered fermions, at different values of 
quark
masses $m_q
= 0.01/a$ ;  $m_q = 0.02/a$.
The results were in this case
\begin{eqnarray*}
a m_q = 0.01\;\; : && T_g = (0.34\pm 0.02)\,{\mbox{Fm}}\quad
G_2 = (0.015 \pm 0.003)\,{\mbox{GeV}}^4 \label{adg3}\\
a m_q = 0.02\;\; : && T_g = (0.29\pm 0.02)\,{\mbox{Fm}}\quad
G_2 = (0.031 \pm 0.005)\,{\mbox{GeV}}^4 \label{adg4}
\end{eqnarray*}

The value of the condensate can be extrapolated 
to the physical quark masses by the techniques of \cite{npgl}
giving $G_2 = (0.022 \pm 0.006) \mbox{GeV}^4$, an order of magnitude
smaller than the quenched value, and in agreement 
with phenomenological determinations   
\cite{6.1,6.2}. It is worth
mentioned, however, that even for full unquenched QCD it is rather
small: $T_g = 0.16$ Fm for quenched $SU(2)$, $0.22$ Fm for
quenched $SU(3)$, and $0.34$ Fm for full QCD with 4 flavours. 

A detailed analysis has also been made 
of the correlators at finite
temperature \cite{lat31,lat32}.
The $O(4)$ invariance is lost at nonzero $T$
and five independent form--factors exist instead of two at $T=0$.

The main result is that magnetic correlators
stay intact in the course of the phase transition,
while
electric correlators have a sharp drop (compare Fig.2 and Fig.3).
Above $T_c$ the electric longitudinal correlators
are very small so that the numerical determination is
affected by large errors.

It is of prime importance to stress, that electric condensate,
defined by the electric parts of (\ref{eq34})
does not play by itself the role of an order parameter
for the confinement--deconfinement
transition and, in particular, stays nonzero above
$T_c$ (since NP part of $D_1$ is nonzero).
It is only the function $D$, which vanishes
at the point of the phase transition and this in its
turn indicates zero string tension (see \ref{sigma})
and hence deconfinement.
This behaviour of correlators at the deconfinement transition is
fully consistent with theoretical
expectations.  More on the field correlators at nonzero
 temperature and deconfinement transition can be found
in section 5.

\subsection{ Relations between correlators}
 Correlators (\ref{2.4})
with different number of $F$'s are not completely independent
quantities.
The dynamical (Schwinger-Dyson type) equations for them
are discussed in \cite{coreq}. In this section
we are concentrating on the identities which must be satisfied
due to the Bianchi identities, $D_\mu \tilde F_{\mu\nu}=0$,
 which hold true for $F_{\mu\nu}$.
The fields
 are nonabelian
and therefore the derivative of the correlator with respect to
$z_{\sigma}$ contains also  the differentiation of the contour
(see references \cite{nast1}-\cite{nast7}):
  \be
  \frac{\partial\Phi(z,z')}{\partial z'_{\gamma}}=i
\Phi(z,z')A_{\gamma} (z') +i (z'-z)_{\rho}
\tilde{I}_{\rho\gamma}(z,z'),
\label{2.15}
\ee
 where the following notation is used:
 $$
  \tilde{I}_{\rho\gamma}(z,z')= \int^1_0 d\alpha\;\alpha\>
\Phi
 (z,z+\alpha(z'-z)) F_{\rho\gamma}(z+\alpha(z'-z))\cdot
 $$
 \be
 \cdot \Phi(z+\alpha(z'-z),z')
 \ee

One easily obtains the relation
between 2--point and 3-point
correlators:
$$
 \varepsilon_{\mu_1\nu_1\sigma\rho}
\frac{dD(z)}{dz^2}=\frac{i}{4}
 \varepsilon_{\mu_2\nu_2\xi\rho}
 \biggl(\lll Tr(F_{\mu_1\nu_1}(z_1)
 \tilde I_{\sigma\xi}(z_1,z_2)
 F_{\mu_2\nu_2}(z_2)\Phi(z_2,z_1))\rrr  - $$
\be
 - \lll Tr(F_{\mu_1\nu_1}(z_1)
 \Phi(z_1,z_2) F_{\mu_2\nu_2}(z_2)
I_{\sigma\xi}(z_2,z_1)\rrr \> \biggr)
\label{2.21}
\ee

The formula (\ref{2.21}) is exact and also holds for
the perturbative parts of the correlators. 
Let us focus our attention
on the NP parts,
assuming their regular behavior at the origin.
One finds \cite{nast5}:
 \be
 \left. \frac{dD(z)}{dz^2}\> \right|_{z=0}=
 \frac{1}{96N_c}\>  f^{abc} \lll F^a_{\mu_1\nu_1}F_{\nu_1\nu_2}^b
 F_{\nu_2\mu_1}^c\rrr .
 \label{2.22}
 \ee

 In a similar way by differentiating any
FC one obtains two pieces:
 "local" one, where the field strength operator is differentiated, and
 "nonlocal" term, where the parallel transporters are
differentiated. Taking an
 exterior derivative the Bianchi identity eliminates the first term,
 while the second ones give rise to a higher FC. Thus one gets an
 infinite set of relations
between FC. For some more complicated
         relations between
 FC, e.g. expressing the second derivative of $D(z)$ the reader
is referred to \cite{ss1}.

\subsection{Nonabelian Stokes theorem}

The important role in the discussed formalism is played by the
nonabelian Stokes theorem. There 
exist many derivations and interpretations of this
result, see \cite{nast1} - \cite{nast7}. One of the simplest ways
\cite{nast7} to derive the Stokes
 theorem for the nonabelian fields is to use the so called
generalized contour gauge \cite{cg} (see also \cite{luk}), where the
vector-potential
satisfies the
following condition $t_{\mu}(x) A_{\mu}(x)=0$ and is expressed as
\be
 A_{\mu}(x)= \int^1_0 d s
 \frac{\partial z_{\rho}(s,x)}{\partial s}
 \frac{\partial z_{\sigma}(s.x)}{\partial x_{\mu}}
F_{\rho\sigma}(z(s,x))
\label{eq76}
 \ee
Here the set of contours $z_{\mu} = z_{\mu}(s,x)$  is such that
 \be
 z_{\mu}(0,x)=x^0_{\mu} \;\; ; \;\; z_{\mu}(1,x)=x_{\mu}\;\; ; \;\;
z_{\mu}(s,x) = z_{\mu}(s'',z(s',x))
 \label{eq77}
 \ee
and $t_{\mu}(x) =
 \left. \frac{\partial z_{\mu} (s,x)}{\partial s}\right|_{s=1}$.

 Consider now the gauge-invariant object --- the Wegner--Wilson 
loop \cite{ww1,ww2} (which we denote as W--loop throughout this review):
 \be
 W(C)=\frac{1}{N_c} \T \> \P \left( i \oint_{C} A_{\mu}(x) dx_{\mu}\right)
 \label{3.10}
 \ee
 where the integration goes over the closed contour $C$. Insertion of
 (\ref{eq76}) into (\ref{3.10}) immediately yields the nonabelian
 Stokes theorem for $W(C)$,
 \be
 W(C)=\frac{1}{N_c} \T \> {\cal P}\exp \left( i \int_S d\sigma_{\nu\mu}(u)
 G_{\nu\mu}(u,x_0) \right)
 \label{3.11}
 \ee
 where $G_{\mu\nu}$ is related to gauge field strength
 $F_{\mu\nu}$ according to (\ref{eq234}) (notice, that
 phase factors along the contours given by $z(s,x)$
equal to unity in the gauge (\ref{eq76})). 
The surface ordering ${\cal P}$ in (\ref{3.11})
is induced by linear ordering ${\mbox{P}}$ in (\ref{3.10}). 
The reference
 point $x_0$ is chosen arbitrarily on the surface $S$
 (for details see \cite{nast31} - \cite{nast7}). Contrary to
the abelian case, the surface $S$ in (\ref{3.11}) must have the
topology of the disk. The inclusion of higher genera is nontrivial
and requires additional holonomies around fundamental cycles
\cite{nast61,nast62,nast63}.

The r.h.s. of (\ref{3.11}) does not depend neither on
 the choice of surface $S$ nor on the choice of the contours
 $z(s,x)$ in (\ref{eq76}) if the nonabelian
 Bianchi identities $D_{\mu}{\tilde F}_{\mu\nu} = 0$ are
 satisfied. To proceed, one has to consider the
average of the W--loop $\lll W(C)\rrr $ over all gauge
field configurations.
In view of (\ref{3.11}) this average can be expressed
through the correlators (\ref{eq234}).
The average $\lll .. \rrr$ and ordering $ {\cal P} $ procedures,
which should be
applied to $W(C)$, are nontrivially related to each other, as
we discuss it in the next section.

\subsection{Stochastic vacuum and
factorization of field components}

\label{sect:fact}

Next we discuss
the picture of the Gaussian vacuum. In this model,
as it was already explained in the introduction,
all field correlators are assumed to factorize
into two-point ones. Thus  one
approximates the complicated ensemble of fields of nonperturbative QCD
by the one satisfying Gaussian distribution.
We will discuss in what follows existing lattice evidences for
such assumption.
It is very important to mention, that the role of stochastic
variables in the approach is played by parallel transported
field strength variables $G_{\mu\nu}$ (\ref{eq234}), not
by the gauge field potentials $A_{\mu}$. While in perturbation theory
one can establish one-to-one correspondence between gauge-covariant
and noncovariant descriptions \cite{mand}, Gaussian dominance
is gauge-invariant statement, and it is not
clear, for example, which property of the
propagators $\lll A..A\rrr$
it corresponds to in covariant gauges.

Whereas in a process with commuting stochastic
variables the Gaussian factorization is uniquely determined, there are
several possibilities for the case of non-commuting variables which we
discuss now briefly.
One way to factorize averages (\ref{2.4}) is
direct matrix factorization
leading to vanishing of higher van
Kampen cumulants \cite{vc1,vc2} and direct exponentiation:
$$
{\lll \T G(1) ...  G(2n) \rrr}_K \sim
(\lll \T G(1)G(2)\rrr\lll \T G(3) G(4)\rrr \dots  \lll \T G(2n-1)
G(2n)\rrr +
$$
\be
+ \mbox{ all path ordered permutations} )
\label{f2}
\ee (we
remind, that all correlators of the odd power vanish in Gaussian model). 
This scheme is however of not predictive power
if we try to evaluate the expectation
value of two or more W--loops with the help of the basic two-point
correlator. In order to go further in this direction we have to make
asumptions on the factorization of the colour compoments of the fields
directly. An obvious choice is to assume 
the following factorization for the
components:
$$
\lll G^{a_1}(1) ... G^{a_{n}}(2n) \rrr_C sim
$$
$$
(\delta^{a_1a_2}\>\delta^{a_3a_4} \dots
\delta^{a_{n-1}a_{n}}\>
 \lll \T G(1)G(2)\rrr \lll \T G(3) G(4) \rrr \dots \lll \T G(2n-1)
G(2n)\rrr +
$$
\be
 + \mbox{all path ordered permutations })
\label{f3}
\ee

However the difference between the matrix van Kampen factorization
${\lll . \rrr}_K$ and the
component factorization
${\lll . \rrr}_C$
is not fully ordered, and hence
it would give only a correction to the two-point cumulant, if we had complete
path ordering. But after applying the nonabelian Stokes theorem the
field strength matrices are only surface ordered and points in the
correlator, which are separated by a large number in surface ordering
can be physically  quite close.

These contributions can
destroy the surface law and give rise to a term proportional to the
3/2-power of the area.
Since surface ordering is much less treated in the literature than path
ordering we discuss the problem in some detail.
Let $\xi(t), \qquad 0\leq t \leq T $ be a non-commuting stochastic
variable describing a centered Gaussian process with correlator:
\be \label{cor1}
\lll \xi(t_1)\xi(t_2)\rrr =\lll \xi(t_2)\xi(t_1)\rrr
= c \cdot\lll \xi^2\rrr \phi(|t_1-t_2|)
\mbox{ with } \phi(0)=1 \ee
where $c$ is some numerical constant of the order of 1 and function
$\phi(x)$ decays with correlation length $a$.

In the expansion of the exponential
$\lll P\exp[\int_0^T \xi(t) \,dt ]\rrr$ we
encounter terms like
$$
\lll P\exp[\int_0^T \xi(t) \,dt ]\rrr = 1 + \int_0^T \,dt_1\int_0^{t_1}
 dt_2\lll \xi(t_1) \xi(t_2)\rrr +
 $$
 $$
+\int_0^{T} dt_1 \int_0^{t_1}dt_2\int_0^{t_2}dt_3\int_0^{t_3}dt_4
\left(
\lll \xi(t_1)\xi(t_2)\rrr\lll \xi(t_3)\xi(t_4)\rrr \right.+
$$
\be
+\left.\lll \xi(t_1)\xi(t_3)\rrr\lll \xi(t_2)\xi(t_4)\rrr + \lll
\xi(t_1)\xi(t_4)\rrr\lll \xi(t_2)\xi(t_3)\rrr \right)+\dots \label{cor2}
\ee For the first nontrivial term we obtain: \be \label{cor3}
\int_0^T \,dt_1\int_0^{t_1} dt_2\lll \xi(t_1) \xi(t_2)\rrr = c \lll
\xi^2\rrr a T \ee
where the value of $c$ depends
on the form of the correlation function. The next terms
crucially depend on the path ordering ( we assume $T\gg a$ ): \be
\label{cor4} \int_0^{T}dt_1
\int_0^{t_1}dt_2\int_0^{t_2}dt_3\int_0^{t_3} dt_4\lll
\xi(t_1)\xi(t_2)\rrr \lll \xi(t_3)\xi(t_4)\rrr  =
 \frac{1}{2} {c}^2 \lll \xi^2\rrr^2 (aT)^2
\ee
\be \label{cor4a}
\int_0^{T}dt_1 \int_0^{t_1}dt_2\int_0^{t_2}dt_3\int_0^{t_3}dt_4
\lll \xi(t_1)\xi(t_3)\rrr \lll \xi(t_2)\xi(t_4)\rrr
= c' \lll \xi^2\rrr^2  a^3 T
\ee
The term of equation (\ref{cor3}) and the term in equation
(\ref{cor4}) contributes to the exponential
$$\exp[\frac{1}{2} c \lll \xi\rrr ^2 a T]$$
whereas the term of equation (\ref{cor4}) is only of order $T$
and therefore modifies the coefficient of $T$ in the exponent by a term
of the order $c' \lll \xi^2\rrr ^2  a^3$.
The details of the factorization can thus always be absorbed
in a modification of the value of $\lll \xi^2\rrr $.

Let us now consider a stochastic variable on a plane $\xi(u,v)$ with
similar properties as the one discussed above (\ref{cor1}) and
investigate the
surface ordered expectation value:
$$ \lll {\cal P}\exp\int_0^Rdu\int_0^T dv  \; \xi(u,v)\rrr , $$
where ${\cal P}$ stands for surface ordering
analogously to (\ref{3.11}).
The first
nontrivial term in the expansion of the exponential,
analogous to (\ref{cor3})
yields a contribution proportional to
$$\lll \xi^2\rrr  a^2 R T $$ independent of
surface ordering. But the terms containig the expectation value of four
values of the stochastic variable can behave quite differently:
Let again $f_1,f_2,f_3,f_4$ denote four ordered
points on the surface. Then the integral
over the ordered expression $$\lll \xi(f_1) \xi(f_2)\rrr
\lll \xi(f_3) \xi(f_4)\rrr $$
yields $\lll \xi^2\rrr^2 a^4 (RT)^2$ which together with the
former term exponentiates and contributes to the area law
$$ \exp[\frac12 c \lll \xi^2\rrr  a^2 R T] ,$$ but the ordering
$$\lll \xi(f_1)
\xi(f_3)\rrr \lll \xi(f_2) \xi(f_4)\rrr $$
is not simply a correction to the first term:
if $f_1$ and $f_3$ are
close but in different rows the points
$f_2, f_4$ can vary over the full row,
thus yielding a factor  of the order
$T$ , and the final contribution will be of the order
$$ \lll \xi^2\rrr^2 a^3 T (RT)$$
and thus not fit into an exponentiation.  So in contrast to the path
ordering where different clusterings, as long as they agreed in the leading
(ordered) term exponentiated to an perimeter (area) law,
in surface ordering we
need a fine tuning
in order to get the latter.

This is a reflection of the rather
artificial feature of the Gaussian model that we have always to treat
correlators where the reference point is fixed from the beginning and
hence might be quite distant from the connection line of the two
external points.

We can  exclude this unphysical contributions
if we specify  the factorization prescription (\ref{f3}) by
the following ordering rule \cite{stfrm}:
if the expectation value is expanded as a power series in the
two-point correlators only the leading powers of the ratio
of the correlation length and the linear dimensions of the loops have
to be taken into account in each order.

It is easy to see that the factorization (\ref{f3}) together with this
rule leads in the case of one loop to the same area law as the matrix
factorization (\ref{f2}).
It will be referred to in the following as the modified
component factorization in the model of the stochastic vacuum.

It is instructive to mention here
a proposal of \cite{n96} based
on a domain picture of the QCD vacuum in which these problems do not
occur and the formal modification of the  component factorization is
justified from a physical point of view.

\newpage

\section{Confinement and field correlators}

\setcounter{equation}{0} \def\theequation{3.\arabic{equation}}

In this chapter we
discuss
 the following question: how confinement is related with
the properties of FC, what is the physical mechanism
behind it and how the cluster expansion operates in the quantitative
description of confinement. In what follows  the role of
the order parameter of confinement  is played by the  averaged
W--loop  \cite{ww1,ww2} (we assume quenched approximation)
\be
 \lll W(C)\rrr = \frac{1}{N_c} \lll \T \>
\P (i\oint_C A_{\mu} dx_{\mu})\rrr = \exp [-V(R)T]
\label{4.1}
 \ee
This is the true order parameter only in case when light quarks are
absent and $V(R)$ is the interquark potential
for (infinitely) heavy static quarks.
      Confinement for light quarks and gluons is much more intricate
phenomenon, which is, however, ensured by the same FC as 
for heavy quarks (see \cite{lq}).

This Section has the following structure: confinement in terms of the cluster 
expansion
is explained in subsection 3.1, corresponding physical aspects
and confinement of the charges in higher representations are reviewed in 3.2,
abelian projection method in terms of field correlators briefly commented
in subsection 3.3, perturbative structure of the cluster expansion is
explained in subsection 3.4, the string profile is discussed in subsection 3.5,
and, finally, subsection 3.6 is devoted to the explanation of the model 
consistency with
the low--energy theorems and related questions.

\subsection{Confinement and cluster expansion}

The nonabelian Stokes theorem and cluster expansion theorem \cite{vc1}
applied to the W--loop
yields the following result:
$$
 \lll W(C)\rrr = \frac{1}{N_c} \left\lll \T \>
\P (i\oint_C A_{\mu}(x)  dx_{\mu})\right\rrr =
$$
$$
 = \frac{1}{N_c}
\left\lll \T \> {\cal P} \exp (i\int_S G_{\mu\nu}(u,
x_0) d\sigma_{\mu\nu}(u))\right\rrr =
$$
\be
= \frac{1}{N_c}\>
 \T \>  \exp\> \left( \sum\limits_{n=1}^{\infty}
\frac{i^n}{n!} \int
 d\sigma(1)..d\sigma(n) \lll\lll  G(1)..G(n)\rrr\rrr
\right)
\label{4.2}
\ee
where we have suppressed the indices, $d\sigma(k) G(k) =
G_{\mu\nu}(u_k, x_0) d\sigma_{\mu\nu}(u_k)$,
and we have used irreducible cumulants.
Due to colour neutrality of the vacuum all
cumulants are proportional to the unit matrix in the colour space
which makes unnecessary colour ordering in the r.h.s. of (\ref{4.2}).

The irreducible correlators $\lll\lll .. \rrr\rrr$ are straightforwardly
extracted from (\ref{4.2})
comparing terms of the same order in $G$. Notice, that there is
only
one ordering in the expression (\ref{4.2}) and this ordering 
is induced by the P--ordering in the original loop (\ref{4.1}).

The most important for us is the bilocal correlator,
which is defined as
$$
\label{bilocal}
\lll G_{\mu_1\nu_1}(x_1)G_{\mu_2\nu_2}(x_2)\rrr = \lll\lll
G_{\mu_1\nu_1}(x_1)G_{\mu_2\nu_2}(x_2)\rrr\rrr +
\lll G_{\mu_1\nu_1}(x_1)\rrr \lll G_{\mu_2\nu_2}(x_2)\rrr
$$

Since $\lll G(k)\rrr \equiv 0$, one can rewrite (\ref{4.2})
identically as
\be
\lll W(C)\rrr = \frac{1}{N_c}\>  \T\>  \exp\> \left( -\frac12
\int\limits_S \int\limits_S d{\sigma}_{\mu\nu}(u)d{\sigma}_{\rho\sigma}(v)
{\Lambda}_{\mu\nu , \rho\sigma}(u,v,S) \right)
\label{4.3}
\ee
where we have defined the global correlator $\Lambda(u,v,S)$ \cite{ss4}
$$
-\frac{1}{2}\> {\Lambda}_{\mu\nu , \rho\sigma}(u,v,S) \equiv 
-\frac12 \>\lll\lll
 G_{\mu\nu}(u,x_0) G_{\rho\sigma}(v,x_0)\rrr\rrr  + $$
 $$
 + \>
\sum\limits_{n=3}^{\infty} \frac{i^n}{n!} \int
d\sigma(3)..d\sigma(n) \biggl[ \lll\lll
G_{\mu\nu}(u,x_0)G_{\rho\sigma}(v,x_0)G(3)..G(n) \rrr\rrr  +
$$
\be
+ \> {\mbox{perm}}(1,2,..n)
\biggr]
\label{4.4}
\ee
and $ {\mbox{perm}}(1,2,..n)$ stands for the sum of terms with different
ordering of $G(u,x_0)=G(1)$ and $G(v,x_0)=G(2)$ with respect to all other
factors $G(k)$. Since $W(C)$ does not depend on the shape of the surface $S$,
the dependence of the global correlator on its arguments is such,
that r.h.s. of (\ref{4.3}) is independent of the choice of $S$
but depends on the contour $C$.
This circumstance explains the termin "global correlator" used for
${\Lambda}(u,v,S)$ in contrast to local correlators which enter in the
r.h.s. of (\ref{4.4}) (note however, the specific meaning of the
definition  "local" -- correlators
$\lll\lll G(1)..G(k)\rrr\rrr$
depend on the points $z_1,..,z_k$ as well as
on the paths, entering
via transporters $\Phi(z,x_0)$).

If one integrates over all but one variables in (\ref{4.3}),
it leads to the effective integrand  
\be
{Q}_{\mu\nu}(u,S) \equiv \frac12 \> \int\limits_{S} d{\sigma}_{\rho\sigma}
(v){\Lambda}_{\mu\nu , \rho\sigma}(u,v, S)
\label{www4}
\ee
In the confining phase one expects for large contours
$C$ the minimal area law of W--loop, which implies
that $Q_{\mu\nu}$ in this limit does not depend on the point
$u$ when $S$ is the minimal area surface and simply coincides with
the string tension $\sigma$, while for the arbitrary surface
one can identify $Q_{\mu\nu}$ as
\be
{Q}_{\mu\nu}(u,S) = P_{\mu\nu}\cdot\sigma
\ee
where $P_{\mu\nu}$ projects onto the minimal
surface. Conversely if $Q_{\mu\nu}$ does not have constant limit
for large $S$ then the area law of W--loop does not hold.

To calculate $Q_{\mu\nu}(u,S)$ one can for simplicity
take $x_0$ in (\ref{4.4}) to coincide with $u$. Then the lowest
order FC in (\ref{4.4}) depends only on two points (as in (\ref{bil})).
 In what follows we concentrate on contributions
to $\sigma$ and therefore take for simplicity a planar contour $C$
with the minimal surface $S$ lying in the plane.

Insertion of (\ref{bil}) into (\ref{4.4})  yields the area law of
W--loop with the string tension $\sigma$
  $$\lll W(C)\rrr = \exp (-\sigma
S_{min}),
$$
 \be
  \sigma^{(2)} =\frac{1}{2}\int d^2 x\>  D(x)
\label{sigma}
\ee
Note that $D_1$ does not enter $\sigma$, but gives rise to the
perimeter  term. It can be clearly seen from the definition 
(\ref{bil}) since $D_1$ by definition contains 
derivatives. 
One concludes therefore, that the presence of nontrivial
nonperturbative contribution to the
function $D(x)$ such, that (\ref{sigma})
is nonzero but finite implies confinement
in Gaussian approximation. The lattice data 
described above support this since $D(x)$ demonstrates 
exponential falloff at large distances and hence 
$$
\sigma \sim G_2 \cdot T_g^2   
$$

Till now we have discussed confinement in terms of the lowest
cumulant -- $D(x)$, which is justified when stochasticity condition
(\ref{3.21}) is fulfilled and $D(x)$ gives a  dominant contribution. Let
us now turn to other  terms in the cluster expansion (\ref{4.2}). It is
clear that the general structure of higher cumulants is much more
complicated than (\ref{bil}), but there always
Kronecker-type terms $D(x_1, x_2, ..., x_n)\prod \delta_{\mu_l \mu_k}$
present
similar to $D(x-y)$ in (\ref{bil}) and other terms containing
derivatives like $D_1(x-y)$. The term $D(x_1,
x_2, ... , x_n)$ contributes to string tension.
 This means that the string
tension in the  general case is a sum,
\be
 \sigma = \sum^{\infty}_{n=2}
\sigma^{(n)}\;, \; \sigma^{(n)} \sim 
\int\lll\lll  G(1) G(2) ... G(n)\rrr\rrr
d\sigma (2) ... d\sigma(n)
\label{4.10}
\ee

If one assumes, that all gluon (irreducible) condensates and
the corresponding correlation lengths
have the same order of magnitude, then  
the expansion in (\ref{4.10}) is in powers of $( \lambda \sqrt{G_2}
T^2_g)$, and if this parameter is small,
\be
\lambda \sqrt{G_2} T^2_g \ll  1, \label{3.21} \ee one gets the limit of
Gaussian stochastic ensemble where the lowest (quadratic in
$G_{\mu\nu}$) cumulant is dominant. The value of $\sqrt{G_2}$ in
(\ref{3.21}) is a typical scale of vacuum fields, which can be
estimated from the gluon condensate, $ \sqrt{G_{2}}\sim
(0.6\> \mbox{GeV})^2$, while $\lambda\sim 1$ takes into account possible
cancellations in the magnitude of the cumulant, and in this way
characterizes the deviation of the vacuum ensemble
from Gaussian one.
 Taking into account, that $\sqrt{G_2}$ in $W(C)$ enters
 on a plane with fixed $(ij)$, i.e. $\sqrt{G_2}\sim\sqrt{\lll G^2_{ij}\rrr }$, 
while
 the gluonic condensate  is a sum $\sum_{ij} \lll G^2_{ij}\rrr $, and
 the numerical value of $T_g$ from lattice calculations $T_g\sim
 0.2-0.3 $ fm, one obtains  as the parameter of cluster expansion
 $\lambda G_2 T_g^2\sim \lambda\cdot 0.1$.

It is of prime importance to stress, that 
the relative contribution of different terms 
in the r.h.s. of the expression (\ref{4.10})
depends on the chosen integration surface.
Since the l.h.s. of (\ref{4.10}) does not depend
on the surface, the lowest cumulant dominance 
is expected to hold for the minimal surface
in Gaussian dominance scenario, while 
contributions from higher terms are to become
important for arbitary nonminimal surface 
in order to cancel 
artificial surface dependence.
In can be said therefore, that the actual value
of the string tension is determined by the total sum
(\ref{4.10}) in any case: the stochastic one (actually governed
by Gaussian average for the minimal surface) and coherent one
(when all terms are
of the same order even for the minimal surface).
 
Summarizing, 
when the stochasticity condition
(\ref{3.21}) holds, the lowest term, $\sigma^{(2)}$, dominates in the
sum; in the general case all terms in the sum (\ref{4.10}) might be
important. Therefore the smaller the correlation length $T_g$ is, 
the better
Gaussian model of confinement works.
It is worth mentioning, that this limit is opposite to
the sum rule limit (sometimes called Leutwyler-Voloshin limit),
where the condensate is assumed to be constant in the position space,
i.e. using the language of MFC $T_g \to \infty$. From this point of view the
analogy between Gaussian dominance in the stochastic picture
and "vacuum insertions" procedure in OPE is
not straightforward.

 \subsection{Physical
mechanism of confinement }

The described formalism is valid both for abelian and nonabelian
theories, and it is interesting whether the area law and nonzero
string tension could be valid also for QED (or
$U(1)$ in the  lattice version \cite{compqed}). To check it let us apply
the operator
$\frac{1}{2}\varepsilon_{\mu\nu \alpha\beta}\frac{\partial}{\partial
x_{\alpha}}$ to both sides of (\ref{bil}) \cite{nast5}.

In the abelian case, when $\Phi$ cancel in (\ref{2.4})
and $F=G$,  one   obtains
\be
\partial_{\alpha}\lll \tilde{F}_{\alpha\beta}(x)F_{\lambda\sigma}{(y)}\rrr =
\varepsilon_{\lambda\sigma\gamma\beta}\partial_{\gamma}D(x-y),
\label{usp}
\ee
where
$\tilde
F_{\alpha\beta}=\frac{1}{2}\varepsilon_{\alpha\beta\mu\nu}F_{\mu\nu}.$

If magnetic monopoles
 are present in abelian theory (e.g. external
Dirac monopoles) with
the current $\tilde{j}_{\mu}$, one has
\be
\partial_{\alpha}\tilde{F}_{\alpha\beta}(x)=\tilde j_{\beta}(x)
\label{4.88}
\ee
In absence of magnetic monopoles (for pure QED) the abelian Bianchi
identity requires that
\be
\partial_{\alpha}\tilde{F}_{\alpha\beta}\equiv  0
\ee
Thus for QED (without magnetic monopoles) the function $D(x)$
vanishes due to (\ref{4.88}) and hence confinement is absent
(compare with (\ref{es2}).

In the compact version of $U(1)$ magnetic monopoles are present 
and the lattice formulation of the method would
predict the confinement regime with nonzero string tension \cite{polyakov1}.
The latter can be connected through $D(x)$ to the correlator of
magnetic monopole currents. Indeed, multiplying both sides of
(\ref{usp}) with $\frac{1}{2}
\varepsilon_{\lambda\sigma\gamma\delta}\frac{\partial}{\partial y_{
\gamma}}$ one obtains
\be
\lll \tilde{j}_{\beta} (x) \tilde{j}_{\delta} (y)\rrr  =
\left( \frac{\partial}{\partial x_{\alpha}} \cdot \frac{\partial}{\partial
y_{\alpha}} \delta_{\beta\delta} - \frac{\partial}{\partial
x_{\beta}} \frac{\partial}{\partial y_{\delta}}\right)  D(x-y)
\label{4.18}
\ee
This identically satisfies monopole current
conservation:
applying $\frac{\partial}{\partial x_{\beta}}$ or
$\frac{\partial}{\partial y_{\delta}}$ to both sides of
(\ref{4.18}) gives
zero. More on MFC in abelian theories can be found in 
\cite{dima2}.

Let us turn now to the nonabelian case, again assuming stochasticity
condition (\ref{3.21}).
Applying as in the Abelian case the operator $\frac{1}{2}
\varepsilon_{\mu\nu\alpha\beta} \frac{\partial}{\partial x_{\alpha}}$
to the r.h.s. of (\ref{bil}), one obtains
the relation given above in (\ref{2.21})
 Especially simple form of
(\ref{2.21}) occurs when tending $x$ to $y$; one obtains
equation (\ref{2.22}), which can be rewritten  as
 \be
  \left.\frac{dD(z)}{dz^2}\right|_{z=0}\sim
\lll E_i^aE_j^bB_k^c\rrr \varepsilon_{ijk} f^{abc}
\label{rua}
\ee
Thus confinement (nonzero $\sigma$ due to nonzero $D$)
is related in nonabelian case with the presence of purely nonabelian correlator.

To see the physical
 meaning of this correlator one can  visualize
magnetic and electric field strength lines (FSL) in the space. Each
magnetic monopole is a source of FSL, whether it is a real object
(classical solution or external object like Dirac monopole) or
lattice artefact.
In nonabelian theory these lines may form branches and e.g. electric
FSL may emit a magnetic FSL at some point, playing the role of
magnetic monopole at this  point. This is what exactly nonzero
triple correlator $\lll \T FFF\rrr $ implies. Note that this could be a
purely quantum effect and no real magnetic monopoles (as classical solutions
or lattice artefacts) are necessary
for this mechanism of confinement.

The formula (\ref{rua}) 
demonstrates, that in strict 
Gaussian approximation $D(z^2)' = 0$,
since the r.h.s. is proportional to the
condensate of the third power.
The extended Gaussian approximation, which keeps only
2-- and 3--point correlators allows to have nonzero $D(z^2)' = 0$.
It does not automatically mean, however, 
that the contribution of the triple (and higher odd order)
condensates themselves to the physical quantities (like string tension)
is comparable with that of 2--point one (see (\ref{4.10}) and
discussion below (\ref{3.21})).
 It should also be noticed, that despite
the first derivative of the function
$D(z)$ vanishes at the origin in Gaussian vacuum, 
the second and higher ones 
of even order do not (see \cite{ss1}).

We conclude this section with discussion of confinement for charges in
higher representations.
Most of the models of NP QCD
are designed to describe confinement of colour
charge and anticharge in the fundamental representation
of the gauge group $SU(3)$, i.e. the area law for the
simplest W--loop and hence linear potential
between static quark and antiquark. At the same time
Gaussian dominance scenario naturally 
describes lattice data on 
the interaction
between static charges in higher $SU(3)$ representations.

We define static potential between sources at the distance $R$
in the given
representation $D$ as:
\be
V_D(R)=-\lim_{T\to\infty}\frac{1}{T} \log \lll W_D(C)\rrr,
\label{eqv1}
\ee
where the W--loop $W_D(C)$ for the rectangular contour $C=R\times T$ in
the "34" plane is given by
\be
\lll W_D(C)\rrr = \left\lll {\T}_D\>{\mbox P}\exp\left(i\int\limits_C
A_{\mu}^a T^a dz_{\mu}\right)\;\right\rrr
\label{eqv2}
\ee

The $SU(3)$ representations $D=3,8,6,15a, 10,27, 24, 15s$
are characterized by $3^2-1=8$ hermitian generators $T^a$ which
satisfy the commutation relations $[T^a, T^b] = i f^{abc} T^c$.
One of the main characteristics of the representation
is an eigenvalue of quadratic Casimir operator ${\cal C}^{(2)}_D $,
which is defined according to ${\cal C}^{(2)}_D = T^a T^a = C_D\cdot \hat1$.
Following the notations from \cite{bali1} we
introduce the Casimir ratio $d_D = C_D / C_F$, where
the fundamental Casimir $C_F = (N_c^2-1)/2N_c$ equals to
$4/3$ for $SU(3)$. The invariant trace is given by $\T_D \hat1 =1$.

The potential (\ref{eqv1}) with the definition (\ref{eqv2}) admits the
following decomposition
 \be
 V_D(R)= d_D V^{(2)}(R) + d^2_D V^{(4)}(R)+...,
 \label{eqv3}
 \ee
where the part denoted by dots contains terms, proportional to
the higher powers
of the quadratic Casimir as well as to higher Casimirs.
It is worth mentioned, that the term $V^{(2)}(R)$ is a sum of the
bilocal correlator contribution and linear Casimir contributions
coming from higher correlators.

The fundamental static potential contains perturbative
Coulomb part,  confining linear and constant
terms
\be
V_D(R) = \sigma_D R - v_D - \frac{e_D}{R}
\label{pppo}
\ee
 The  Coulomb part is now known up to two loops \cite{psr1,psr2}
and is proportional to $C_D$.
The "Casimir scaling hypothesis"
\cite{amb}, see also \cite{go1,go2,go3,go4} declares, that
the confinement potential is also proportional to the
first power of the quadratic Casimir $C_D$, i.e.
all terms in the r.h.s. of (\ref{eqv3}) are much smaller than
the first one. In particular, for the string tensions
one should get $\sigma_D/\sigma_F = d_D$.

This scaling law is in perfect agreement with the results
found in \cite{bali1} (see also  \cite{bali3}).
It is easy to see that Gaussian cumulant in
(\ref{eqv2}) is expressed through
$C_D$ and representation--independent averages as
\be
\T_D\lll F(1) F(2)\rrr =
\frac{C_D}{N_c^2 -1}\> \lll F^a(1)
F^a(2)\rrr =
\frac{d_D}{2N_c}\> \lll F^a(1)
F^a(2)\rrr,
\label{eq4}
\ee
so Gaussian approximation satisfies "Casimir scaling law"
exactly. This fact does not depend
on the actual profile of the potential.
It could happen, that the linear potential observed
in \cite{bali1} is just some kind of
intermediate distance characteristics and changes the profile
at larger $R$ (as it actually should happen
due to the screening of
the static sources by dynamical gluons from the vacuum)
The coordinate dependence
of the potential
is not directly related to the Casimir scaling,
 and can be analized
at the distances which are small enough
not to be affected by the screening effects
(see discussion in \cite{ss51,ss52,ss53}).

The quantitative analysis \cite{ss52} of the data from \cite{bali1}
shows, that scaling holds with an accuracy of a few percents.
The Fig. 4, taken from the paper \cite{bali1}
shows the potential between static sources in the different
representations of the gauge group $SU(3)$. Lines on this figure 
correspond to the fit of the fundamental static potential 
multiplied by the
corresponding ratio of the Casimir factors. The picture demonstrates nice
agreement with the Casimir scaling hypothesis.
It is worth mentioning that most of the 
microscopic models used in NP QCD
(e.g. instanton model, center vortex model, abelian dominance
hypothesis) encounter
difficulties in explanation of Casimir scaling \cite{ss51,ss53}.

In the MVC the Casimir scaling has
two important features. First, it is the direct consequence of the
Gaussian dominance hypothesis since bilocal correlator provides
the exact Casimir scaling. On the other hand, it implies the
cancellations of $C_D^2$ -- proportional terms and higher ones in
the cluster expansion of (\ref{eqv2}). Physically, it means the
picture of the vacuum, made of relatively small colour dipoles
with weak interactions between them. One can imagine two possible
scenario. According to the first one, Casimir scaling is just a
consequence of Gaussian dominance. It this case any higher
cumulant contributes to physical quantities much less than the
Gaussian one due to dynamical reasons.
 There is also the second
possibility, when each higher term in the expansion (\ref{eqv2})
is not small, but their sum demonstrates strong cancellations
of Casimir scaling violating terms. These pictures are in close
 correspondence to the stochastic versus coherent vacuum
scenario, discussed above. The clarification of the
situation in future lattice 
measurements is important to understand the NP structure of QCD
vacuum in gauge-invariant terms.

\subsection{Perturbative analysis of correlators in gluodynamics}

Perturbative expansion of FC has specific features
due to the presence of parallel transporters in the
definition of FC, which should also be expanded
and generate new class of diagrams. In this respect FC are similar
to W--loops. At the same time since FC contains a product of
operators one expects at small distances singular contributions as in the
OPE coefficient functions.
Let us briefly describe the
renormalization procedure for W--loops. It
was studied both for smooth and cusped contours \cite{renw1,renw2}. In the first
case the result is:
\be W(C) = Z\cdot
W_{ren}(C) \label{ren}
\ee
where the (divergent) $Z$-factor contains linear
divergencies arising from the integrations over the contour
\be
Z\sim {\exp}\left({c\>\int\limits_{C}
dx_{\mu}\int\limits_{C}
dy_{\nu} D_{\mu\nu}(x-y)}\right)
 \sim {\exp}\left({c\> \frac{L}{a}}\right)
\ee
where $L$ is the length of the contour,
$a$ -- ultraviolet cutoff and $c$ --
numerical constant not important for us here.
All logarithmic divergencies are absorbed into
the renormalized charge
$g_{ren}(\mu)$ defined at the corresponding dynamical scale $\mu$.
It is worth noting that a single diagram in the perturbative
expansion of the r.h.s. of (\ref{ren}) typically gives contributions to
both $Z$-factor and $W_{ren}(C)$, i.e. contain linearly and
logarithmically divergent parts. Hence the factorization in the
r.h.s. of (\ref{ren}) has double-faced nature - it is factorization
of the diagrams together with the separation of contributions,
coming from dangerous integration  regions (i.e. linearly divergent terms)
for each diagram itself. As a result contour divergencies can be separated
as a common  $Z$-factor, which physically renormalizes the bare mass of
the test particle, moving along the contour, which forms W--loop.
Therefore the W--loop can be calculated or measured on the lattice only
modulo divergent $Z$-factor.
However
if one extracts the physical quantities from the W--loop average
(e.g. by forming the Creutz ratio)
then only logarithmic charge renormalization
causes an observable effect.

If the contour possesses cusps (and selfintersections) situation
becomes more complicated. Namely, each cusp leads to its
own local $Z$ -- factor, depending on the cusp angle (but not on
other characteristics of the contour).
The expression (\ref{ren}) for the loop with $k$ cusps having
angles ${\gamma}_i$ is to be modified
in the following way \cite{renw1,renw2}:
\be W(C) = \prod\limits_{i=1}^{k} Z_i\>({\gamma}_i)\cdot
W_{ren}(C) \label{ren1}
\ee
The mixing between linear and logarithmic divergencies, mentioned
above has some specific features for the loops with cusps.
Namely, each cusp in addition to the linearly divergent factors,
as in the smooth contour case
introduces
logarithmically  divergent terms which have nothing to do with the
coupling constant renormalization and can be subtracted by their
own $Z({\gamma}_i)$--factors. There are two types of terms of
that kind, those, depending on the cusp angle and others which do
not depend on it. The latter play important role in
the perturbative expansion of FC.
To see this, note, that FC can
be easily
 obtained from the W--loop functional.
Indeed, differentiating (\ref{4.2})
with respect to
${\delta}{\sigma}_{\mu\nu}(x_i)$
and assuming that the contour
$C$ connects the points $u, v$ with the
reference point $x_0$ along the straight lines,
one finds for the bilocal
correlator:
\be
\frac{1}{N_c} \T  \lll\lll G_{\mu\nu}(u,x_0)
G_{\rho\sigma}(v,x_0)
\rrr\rrr
= \frac {{\delta}^2 \lll W(\tilde C)\rrr }{\delta
{\sigma}_{\mu\nu}(u) \delta
{\sigma}_{\rho\sigma}(v)}
\label{qu}
\ee
and analogous expressions for higher cumulants.
Note, that
this contour generally
has at least two cusps (typically even four, if the reference point
$x_0$ does not coincide with one of the correlator arguments).

Comparing (\ref{ren1}) and (\ref{qu}) one concludes, that
the derivatives in the r.h.s. of (\ref{qu}) act on the renormalized
part of the W--loop as well as on the $Z$ -- factors.
As a result a mixing between different terms appears
and contour divergencies cannot be separated in a
simple way in the case of FC,
and the perturbative
behaviour of FC is
controlled by both ultraviolet
and contour logarithmic divergencies.

Direct perturbative analysis of the simplest nontrivial bilocal
FC \cite{ej,sh1} shows, that this is indeed the case.
In a particular case where the reference point coincides with one of
correlator arguments the perturbative expressions for the functions
$D$ and $D_1$  at the next-to-leading order reads \cite{ej}:
\be
D^{(1)}(z^2)  =  g^2 \frac{N_c^2-1}{2}\frac{1}{\pi^2
z^4}\frac{\alpha_s}{\pi} \biggl[-\frac{1}{4}L+\frac{3}{8}\biggr]
\label{eq:2.3}
\ee
\be
D_1^{(1)}(z^2)  = g^2 \frac{N_c^2-1}{2}\frac{1}{\pi^2
z^4}\frac{\alpha_s}{\pi}\cdot
\biggl[\biggl(\frac{\beta_1}{2N_c}-\frac{1}{4}\biggr)L+\frac{\beta_1}{3N_c}+
\frac{29}{24}+\frac{\pi^2}{3}\biggr]  \label{eq:2.4}
\ee
with $L=\mbox{log}(e^{2\gamma}\mu^2 z^2/4)$, $\mu$ being a renormalization
scale in the $\overline{ MS}$ scheme and the first coefficient of
the $\beta$-function $\beta_1=(11N_c-2f)/6$.

Note, that at the tree level $D$-function receives zero contribution
while $D_1$ is:
\be
D_1^{(0)}(z^2) =
\frac{N_c^2-1}{2N_c}\,\frac{g^2}{\pi^2 z^4}
 = \frac{16}{3}\> \frac{{\alpha}_s}{\pi}
\> \frac{1}{z^4}
\ee

Since the string tension is an integral
over $D(x)$ (\ref{eq:2.3}) tells that the ${\cal O}(g^4)$ perturbative
contribution to $D(x)$ produces nonzero string tension, which
physically has no sense and should be cancelled by other
terms of the cluster expansion.
Such cancellation was indeed shown to occur
\cite{ss4} in perturbation theory, so string tension is exactly zero at any
finite
 order of the perturbation theory in the gauge--invariant 
formalism considered, i.e. if the latter
is developed in terms of FC.

\subsection{The string profile}

It was already mentioned above that the main phenomenon behind the
physics of confinement is the formation of the string   between the
colored sources. It is this string  which mediates attraction between
quark and antiquark with the force of roughly 15 tons.

An immediate question arises: what does the string consist of? Or in
other words; what is the distribution of fields inside the string?

Before answering this question, two specifications are needed. First,
speaking about field in the string, one should always compare it  to
the field in the vacuum, since only this difference makes the
string. Second, in the QCD vacuum, which is stochastic by necessity
as discussed above, the gauge invariant field distribution can be
measured only with the help of field correlators.

It is instructive to distinguish two general types of 
correlators in this respect:

{\bf i)} connected correlators -- correlators of colored objects, which we discussed above,
where the trace operation appears only once (if at all)
$$
  \lll\lll \T G_{\mu\nu}(u,x_0)
G_{\rho\sigma}(v,x_0)
\rrr\rrr
$$

{\bf ii)} disconnected correlators -- 
correlators of two or more white objects, like
$$
\lll\lll  \T F^2_{\mu\nu}(x)\> \T F^2_{\lambda\gamma}(y)\rrr\rrr
$$
In nonabelian theory, unlike in QED, those two correlators correspond
to two very different probes, which can be made with fields in
the string. In the first case one measures the color field
distribution as it is, in the second one measures the "interaction"
between two white objects: the string and the neutral plaquette,
which can exchange only white objects -- glueballs in case of
gluodynamics. These approaches are complementary
to each other and  in both cases one gets some gauge-invariant 
characteristics of the problem. 
In what follows we will compare the results of 
analysis performed  with these two kinds of probes. 

We start with the case of the connected correlators.
Using Monte-Carlo (MC) technique,
in \cite{lat41,lat42,lat43} the study was made of
 the spatial distribution of the components of
the field strength tensor in presence of a static $q\bar{q}$ pair.
 Following \cite{lat42}, we define:
\be
\rho_{\mu\nu} = \frac{\lll
\T(W\cdot \Phi \cdot P_{\mu\nu}(x_{\Vert},x_{\bot}){\Phi}^+) \rrr}{\lll \T\>
W \rrr}-1
\label{ppp}
\ee
 where $W$ is a W--loop, $\Phi$ is a phase factor
 and
$P_{\mu\nu}$ is the  plaquette,  oriented in order to give the
desired component  of the field.\footnote{Indices $\mu , \nu$ in
(\ref{ppp})
are not usual tensor indices but numerical arguments, informing about
the chosen direction of the plaquette.}
 The coordinates $x_{\Vert},
x_{\bot}$ measure  the distance from the edge of the 
W--loop and from the plane defined by the loop, respectively.
 In
the naive continuum limit $a \to 0$,
 \be
  \rho_{\mu\nu}\simeq a^2
\lll F_{\mu\nu} \rrr_{q\bar{q}}
\ee
where $\lll F_{q\bar q}
\rrr $ is the average value of the complicated Wilson contour
with plaquette $P_{\mu\nu}$ reduced to $F_{\mu\nu}a^2$.  In
 \cite{lat43} a $16^4$ lattice was used, taking a $8\times 8$ W--loop
and $\beta = 2.50$, which is inside the scaling window for the
fields.  Moving the plaquette in and outside the plane defined by the
W--loop, one obtains a map of the spatial structure of the field
as a function of $x_{\Vert}$ and $x_{\bot}$.  Using a controlled
cooling technique, one
eliminates the short-range fluctuations \cite{cool}.
The long--range
non--perturbative effects survive longer to the cooling procedure,
showing a plateau of 10--14 cooling steps, while the error becomes
smaller. A similar behaviour has been observed for the string
tension and the magnitude of string tension at the  plateau coincides
with the full uncooled value to a good approximation, which confirms
the physical significance of string measurements, done in
\cite{lat41,lat42,lat43}.

 The cooling technique allows  to disentangle the signal from
the quantum noise with a relatively small statistics. The general
patterns of the field configurations are briefly resumed in the
following figures, while the complete results for the field--strength
tensor are contained in Tables of \cite{lat42,lat43}.  Fig. 5 represent a
map of the spatial behaviour of the longitudinal component
of the CE field.
From Figs. 5, it is seen
that the parallel electric field is squeezed in flux tube.


We denote direction along the $q\bar{q}$ axis $x_{\Vert} = x_1$,
while that of
 $x_{\bot} = x_2$, and the Euclidean temporal axis is $x_4$. All
the construction is taken at a fixed value of $x_3$.

Using the non-abelian Stokes theorem and the cluster expansion
(see chapter 3) for
$\rho_{\mu\nu}$  one has
\be
 \rho_{\mu\nu}(x_1,x_2,x_4)= a^2\int
d\sigma_{14}(x'_1,x'_4) \Lambda_{\mu\nu}
\label{8.8}
\ee
  where
   \be
\Lambda_{\mu\nu} =\frac{1}{N_c}\T \langle E_1(x'_1,0,x'_4)\Phi
F_{\mu\nu}(x_1,x_2,x_4) \Phi^+ \rangle +...,
\ee
$\Phi$ is the parallel transporter from the point $(x'_1,0,
x'_4)$ to $(x_1,x_2,x_4)$, and dots imply contribution of
higher order cumulants containing additional powers of $E_1$.

We shall keep throughout this section only the lowest cumulants
(containing lowest power of $E_1$) and   compare our prediction
with the MC data of previous sections. The  bilocal correlator
$\Lambda_{\mu\nu}$  can be expressed in terms of two independent
Lorentz scalar functions $D((x_{\mu}-x'_{\mu})^2)$,
$D_1((x_{\mu}-x'_{\mu})^2)$ (see chapter 2)
\be
\Lambda_{14}=D+D_1+(h^2_1+h^2_4)\frac{dD_1}{dh^2}
\ee
\be
\Lambda_{24}=(h_1h_2)\frac{dD_1}{dh^2}~~,~~~
\Lambda_{34}=(h_1h_3)\frac{dD_1}{dh^2}
\ee
\be
\Lambda_{23}\equiv 0,~~
\Lambda_{13}=(h_3h_4)\frac{dD_1}{dh^2}~;~~
\Lambda_{12}=(h_2h_4)\frac{dD_1}{dh^2}
\ee
Here $h_{\mu}=(x-x')_{\mu}$.

\noindent
Since all construction in Fig. 5 is at $x_3=x'_3=0$ we have $h_3\equiv 0$
and hence
\be
\rho_{23}=\rho^c_{34}=\rho^c_{13} \equiv 0
\ee
The only nonzero components are $\Lambda_{14}, \Lambda_{24}$ and
$\Lambda_{12}$.

When $x_4=0$ ( and this is where measurements of $\rho^c_{12}$ have been
done), $\rho_{12}$ vanishes  because of antisymmetry
 of $\Lambda_{12}$ in $h_4$.
Hence only $\rho_{14} $ and $\rho_{24}$ are nonzero, and only
those have been measured to be nonzero.

On the Fig.5 (a,b,c) 
the field distributions $(\rho_{14}(x_1,x_2))^2$ are shown
for different quark separations, $ R= T_g, 5T_g$ and $15T_g$.
 The string profile, i.e. $(\rho_{14}(x_2))^2$ distribution in the middle of the string, for these quark 
  separations is shown in Fig.5 (d). One can see a clear string-like  formation with the  width of $2.2T_g$.
 The mean value of the CE field in the middle of string does not depend on $R$ for 
$R$ greater then $5T_g$.

To make comparison with data more  quantitative,
one can exploit the  exponential form of $D,\> D_1$ \cite{lat2}
\be
D_1(h^2)= D_1(0)\> \exp(-\delta_1 |h|);\;\; D(h^2) = D(0)\> \exp (-\delta |h|)
\label{8.15}
\ee
$$D_1(0) \approx \frac{1}{3} D(0);~~ \delta_1\approx \delta $$
Inserting this into (\ref{8.8}), we have
\be
\rho_{14}(x_1,x_2;0)= a^2\int^R_0
dx'_1\int\limits^{\frac{T}{2}}_{-\frac{T}{2}} dx'_4
\left[D(0)+D_1(0)-\frac{(h^2_4+h^2_1)}{2h}D_1(0)\right]e^{-\delta h}
\label{8.16}
\ee
with
$$h_4=-x'_4~,~~h_1=x_1-x_1'~~,~~ h^2=h^2_4+h^2_1+x^2_2;$$

\noindent
For $\rho_{24}$  one obtains
a similar expression, which vanishes for $x_2=0$ and in the middle of
the string, which exactly corresponds to the data of \cite{lat31}.

Finally, we can make a detailed comparison of the prediction for
$\rho_{14}$ in (\ref{8.16}) with present lattice data. One
obtains a simple analytic result  for $\rho_{14}(x_2\equiv
x_{\bot})$ in case of a very long string.  The transverse shape
measured at the middle is given
by
\be
\rho_{14}=\frac{2\pi a^2}{\delta^2}\left[D(0)(1+\delta x_2) -
D_1(0)\frac{1}{2} (\delta x_2)^2\right]e^{-\delta x_2}
\label{8.18}
\ee
 $\rho_{14}$ was calculated  as a function of $x_1,x_2$ from
 (\ref{8.16}) keeping $D_1(0)=\frac{1}{3} D(0)$.
 One finds $
\delta^{-1} \approx 0.2 \>{\mbox{Fm}}\>$,
which is in good agreement with \cite{lat2}.

\noindent
The results allow to predict all curves for other values of
$x_{\Vert}$ and $x_{\bot}$: the agreement with the numerical results
is very satisfactory \cite{lat31,lat32}.

The asymptotics of the string profile at large $x_{\bot}$ is
shown to be exponential, see eq.
 (\ref{8.18}), just as the asymptotics of $D,D_1$,
measured in
 \cite{lat11,lat12}.  This in contrast to the behaviour inside the string, where
the Gaussian--like  flattening is observed. This effect is connected
 to
  the smearing
effect due to integration in (\ref{8.16}), yielding polynomial
factors in (\ref{8.18}). The radius of the string is close to
$\delta^{-1}=T_g$ and hence gluon correlation length $T_g$  has
another physical meaning -- the thickness of the confining string.

One can define higher irreducible correlator $\bar{\gamma}^c$
(the superscript $\phantom{1}^c$ is written for $connected$, so the
trace appears in (\ref{8.7}) only once) as
follows
\be
\bar{\gamma}^c
 \approx
a^4[\langle FF'\rangle_{q\bar{q}}-\langle FF'\rangle_0]
\label{8.7} \ee From (\ref{8.7}) it is clear, that
$\bar{\gamma}^c$ contains only  double plaquette correlations.
Most of the lattice data for  $\bar{\gamma}^c$ are compatible with
zero net effect, within two standard deviations (see
\cite{lat32}
and recent analysis in \cite{bbv}).

The same line of reasoning can be elaborated in the case of $qqq$
states, i.e. baryons. The reader is referred to the paper \cite{kuzm},
where the corresponding field profiles of the string were obtained.
An interesting difference from the meson case considered so far 
is the vanishing of the average value of the CE field at the
point of the string junction in the baryon, occuring due to symmetry
reasons. This fact does not mean of course the vanishing of higher order 
correlators, measuring the fluctuations of the fields around the mean 
value (see \cite{kuzm}).

For the double correlator $\bar{\gamma}^c_{\mu\nu}$  one can use
the non-abelian Stokes theorem and the cluster expansion, to
connect
 $\bar \gamma^c$ to the quartic correlator
 $G_4$ and hence to estimate it from
the lattice   data.
Representing the
coordinate dependence of $G_4$ as an exponent similarly to $G_2
\equiv D(h^2)$ with the same $\delta$, one gets the dimensionless
ratio as 
$$\frac{G_4(0)}{(G_2(0))^2}=\frac{\bar{\gamma}^c(0)}
{(\rho(0))^2}\approx 2-3$$ From the point of view of the cluster
expansion convergence, the expansion parameter is of the order
$G_2(0) T_g^4\ll 1$, and one finds that $G_4(0) T_g^8\sim
(2-3)\cdot (G_2(0)T_g^4)^2$ in agreement with the idea of the
bilocal correlator dominance.

We turn now to the case of the disconnected correlators.
The advantage of this probe is that it measures 
the field configurations induced by a
static quark--antiquark pair which are gauge invariant by itself, i.e. 
in the simplest case the traces of the
squares of the CE and CM field components:
 $\sum_a \left( E_i^a \right)^2 ,\; \sum_a \left( B_i^a \right)^2$. 
These expressions are
directly related to physical quantities as for instance  the energy density. 
We are interested in the difference between the squared field
strength of the situation with a static quark-antiquark pair and the
vacuum expectation value without such pair.  This
difference which we denote by $\Delta F^2_{\alpha\beta}\:(x)$ is given by:
\begin{equation}\label{Def F}
\Delta F^2_{\alpha\beta}\:(x)\; \equiv \; \frac{4}{a^4}
\frac{\lll \T W[{\cal C}]\: \T P_{\alpha\beta}(x)\rrr-\lll \T W[{\cal
C}]\rrr \lll\T P_{\alpha\beta}(x)\rrr }{\lll \T W[ {\cal C}]\rrr }.
\end{equation}

Here $\T P_{\alpha\beta}$ is the W--loop of a small square in the
$\alpha-\beta$ plane. From it we obtain the desired gauge invariant squared field
strengths:
\be
{\rm \T} \, P_{\alpha\beta}(x)\; = \;
N_c-\frac{1}{4} a^4\: \sum_a
F_{\alpha\beta}^a(x)F_{\alpha\beta}^a(x)\; +\; {\cal
O}(a^8)\label{edef}
\ee

We see that we have to deal here with a more complex problem than in
the previous case,
namely the evaluation of the expectation value of two loops. 
Therefore the factorization hypothesis of the Gaussian model is tested
here much more severely: vacuum expectation values contain at least four
gluon field tensors which have to be reduced to two-point correlators
by the factorization assumption discussed above.

We again apply the non-abelian Stokes theorem to the line integrals in
the loops and transform then in surface integrals as discussed before.
Expanding the exponentials in eq.(\ref{Def F}) we
obtain:
\begin{eqnarray}\label{Ber F 1}
\Delta
F^2_{\alpha\beta}\:(x) &=& \frac{4}{a^4} \frac{1}{\lll W\rrr} \Bigg(
\sum_{n=1}^\infty (-i)^n \stackrel{\rm surface ordered}{\int \cdots
\int} d\sigma_{\mu_1\nu_1}^W\cdots d\sigma_{\mu_n\nu_n}^W\: \T\left[
{ t}^{a_1}\cdots {t}^{a_n} \right] \times \nonumber \\&&\hspace{-1cm}
\int\int d\sigma_{\mu\nu}^P d\sigma_{\rho\sigma}^P \times
\frac{(-i)^2}{2!}\T \left[ {t}^a { t}^b \right] \lll
F_{\mu_1\nu_1}^{a_1} \cdots F_{\mu_n\nu_n}^{a_n}\:
F_{\mu\nu}^{a}F_{\rho\sigma}^{b}\rrr \; - \nonumber \\&&\hspace{-1cm}\sum_{n=1}^\infty
(-i)^n \stackrel{\rm surface-ordered}{\int \cdots \int}
d\sigma_{\mu_1\nu_1}^W\cdots d\sigma_{\mu_n\nu_n}^W\: \T\left[ {
t}^{a_1}\cdots { t}^{a_n} \right] \: \int\int d\sigma_{\mu\nu}^P
d\sigma_{\rho\sigma}^P \times \nonumber \\&&\hspace{-1cm}\frac{(-i)^2}{2!}\T
\left[ { t}^a { t}^b \right] \lll  F_{\mu_1\nu_1}^{a_1} \cdots
F_{\mu_n\nu_n}^{a_n}\rrr \; \lll  F_{\mu\nu}^{a}F_{\rho\sigma}^{b}\rrr
\Bigg)\end{eqnarray}

The surfaces $S_W$ and $S_P$ with the elements $d\sigma^W $ and $d\sigma^P$ are bordered by
the loop of the static quarks and the plaquette respectively
and contain both the reference point $y$, the choice of these surfaces
will be discussed later. The indices of the
surface elements are restricted to $\mu < \nu$ .

Using the factorization prescription given in the previous section  in
 and subtracting out the terms 
where the fields on $S_W$ and $S_P$ factorize 
separately we obtain:
\begin{eqnarray}\label{Ber F 2}
&&\hspace{-7mm}\Delta F^2_{\alpha\beta}\:(x) = \frac{4}{a^4} \frac{1}{\lll W\rrr} \Bigg(
\sum_{n=1}^\infty (-i)^{2n} \stackrel{\rm surface ordered}{\int \cdots
\int} d\sigma_{\mu_1\nu_1}^W\cdots d\sigma_{\mu_{2n}\nu_{2n}}^W \times
 \\
&& \T \left[ { t}^{a_1}\cdots { t}^{a_{2n}} \right] \;
\int\int d\sigma_{\mu\nu}^P d\sigma_{\rho\sigma}^P
\frac{(-i)^2}{2!2} \Bigg[  \sum_{\rm pairs}^{\alpha_1,\beta_1\cdots
\alpha_n,\beta_n} \lll  F_{\mu_{\alpha_1}\nu_{\alpha_1}}^{a_{\alpha_1}}
F_{\mu_{\beta_1}\nu_{\beta_1}}^{a_{\beta_1}}\rrr \cdots \nn \\
&&\lll 
F_{\mu_{\alpha_{n-1}}\nu_{\alpha_{n-1}}}^{a_{\alpha_{n-1}}}
F_{\mu_{\beta_{n-1}}\nu_{\beta_{n-1}}}^{a_{\beta_{n-1}}}\rrr \times \lll 
F_{\mu_{\alpha_n}\nu_{\alpha_n}}^{a_{\alpha_n}} F_{\mu\nu}^a\rrr \; \lll 
F_{\mu_{\beta_n}\nu_{\beta_n}}^{a_{\beta_n}} F_{\rho\sigma}^a\rrr \Bigg]
\Bigg).\nn \label{Herl f 1} \end{eqnarray}
Of the $2n$
field strengths running over the surface $S_W$ two are
correlated with a $F_{\mu\nu}^C$ on the pyramid of the plaquette and
the other $(2n-2)$ among one another.  We now use the prescription
developed in the previous section and
neglect (in any order of the expansion of the exponentials) all
terms which are not fully ordered, i.e. which are suppressed by at
least a factor $T_g/L$. For all fully ordered expressions the traces in
(\ref{Ber F 2}) are identical and one
obtains finally after a lengthy calculation \cite{stfrm}:

\begin{equation}\label{End}
\Delta F^2_{\alpha\beta}\:(x) \; = \; \frac{1}{a^4}
\frac{1}{2N_c(N_c^2-1)} \left[ \int \int d\sigma_{\mu\nu}^W
d\sigma_{\rho\sigma}^P \lll F_{\mu\nu}^a F_{\rho\sigma}^a \rrr \right]^2\;
\end{equation}

We note that in the integral one surface element is
connected to the surface $S_W$ (indicated by the index $W$) and one
to the plaquette oriented in $\alpha,\beta$-direction (indicated by
$P$)
As noted above the choice of the surfaces is somewhat arbitrary and
depends on the way by which we transport colour flux from the point $x$ to
the reference point $y$. Choosing a straight line it is reasonable to
put the reference point between the plaquette and the W-loop. In this
case the surfaces are the sloping sides of pyramides with the reference
point as apex.  It was checked in \cite{DNR95}, \cite{stfrm}, that different choices of the
surfaces leads only to moderate changes in the results. A reasonable
strategy seems to be to choose the reference point in such a way that
the total resulting surface is minimal.

For the evaluation of (\ref{End})  we refer to the paper \cite{DNR95}, \cite{stfrm}
and quote only the main results in below.
First of all, it can be seen by symmetry arguments that for that case the difference
of the square of the magnetic field strengths, i.e.~the plaquette
$P_{\alpha\beta}(x)$ with no time ($x_4$) component, vanishes
identically. This means that the CM background field is not
affected by the static color charges.

The electric field
perpendicular to the W--loop is also practically not
affected but only the difference of the squared electric field parallel
to the loop is.
In both cases the squared electric field difference is
negative, i.e.~the presence of the static source diminishes the vacuum
fluctuations

In Figs.6,7 we display the value of the
squared field strength parallel to the spatial loop extension $\Delta
F^2_{34}\:(x)=-\lll E_z^2(x_3,r)\rrr_{\rm q \bar{q}-vacuum}$ as a function
of the coordinates $r= \sqrt{x_1^2+x_2^2}$ and $x_3$ (the results are
rotational invariant around the $x_3$-axes) for different spatial
separations $R_W$ (notice, that $a=0.35$ fm on these pictures is a chosen scale of 
distance analogous to the correlation length $T_g$ and has nothing to do with  
the lattice spacing $a$ used above).

The
squared field strength reaches its saturation value $\Delta
F^2_{34}\:(x)$ which approximately equals 
to $ 14 {\rm {MeV}/ {Fm^3}}$ for a spatial
extension
of the W-loop $R_W\approx 4T_g$. The transversal
extension, defined by
\begin{equation}
{\bar r} \equiv \sqrt{\frac{\int dr \: r \: r^2 \: \Delta
F^2_{34}\:(0,r)}{\int dr \: r \: \Delta
F^2_{34}\:(0,r)}},
\end{equation}
is practically independent of $R_W$
and about $1.8$ times the correlation length $T_g$. 
As
mentioned above, the quantity $\Delta F^2_{34}\:(x)$ is the difference
between the squared field strength in the presence of a 
static $\bar q q$-pair and the non-perturbative vacuum field strength. 
One can construct several quantities of interest from it.
For example, the figure 10
from the reference \cite{DNR95} 
displays the unsubtracted squared field tensor of the
color source in units of the vacuum gluon condensate
$\lll F^2\rrr$, while the figures 11--13 
from the same reference shows the squared 
field strength including the
perturbative, i.e. Coulomb contribution.

The result is therefore very satisfying for a physical
intuition: a CE flux tube is formed between the
quark-antiquark pair whose energy content increases linearly with the
separation of the pair. One may even compare the energy content of the
flux tube formed in that way with the quark antiquark potential
calculated with the help of a single W-loop. In order to do that we
discuss two low energy theorems in the next subsection.

The analysis of higher cumulants contributions for the disconnected 
correlators was performed
in \cite{d3}.  It was shown, how quartic
cumulant can modify the profile of the string and, in particular, lead to
(small) nonzero components of the magnetic fields averages.

The pictures displayed on the Fig.5 and Figs.6,7 
are rather similar, despite they were obtained 
by use of different probes, and both are in good
agreement with the lattice results for the string profile.
We found it instructive to demonstrate how FC
formalism works in these two different settings: in the first case 
the mean value of the CE field in the presence of the static 
pair was calculated in the Gaussian approximation and its squared average was
shown on the Fig.5; in the second case the correlation of two 
gauge-invariant object was taken and reduced in the Gaussian approximation  
to the integral of the square of the two-point correlator again. 
From general point of view the typical correlation length, characterizing
the decrease of the fields in the case of connected correlators 
is related to the mass of the gluelumps (and to $T_g^{-1}$ in the Gaussian
approximation), while that of the second case represent somehow the glueball
mass, since it corresponds to the interaction of two white objects.

Summarizing, 
all the results discussed so far, together with the arguments on Casimir scaling
 supports the  fundamentality of the
bilocal correlator, which defines the nonperturbative dynamics of
confinement. On the contrary, there are no data known to
the authors up to now, which would indicate a significant
role played by the highest field strength 
correlators.

\subsection{Low energy theorems and consistency of the model}

In this subsection we consider two low energy theorems and apply them
to the results of the previous sections in order to check the
consistency of the model. We follow here closely the line of reasoning
given in \cite{DNR95}.
Let us consider the
vacuum expectation value of a W--loop:
\be
\lll W[C]\rrr = \frac{1}{N} \lll \T \P [i\int_C {A}_\mu d x_\mu]\rrr
\ee
 Here $C$ denotes a closed loop.
 Differentiating $\log \lll W[C]\rrr$
with respect to $-\frac{1}{2g^2}$ one obtains:
$$
 8\pi
\al_{s0}^2 \frac{\partial\log[\, \lll W[C]\rrr \, ]}{\partial
  \al_{s0}} =
\frac{1}{\lll W[C]\rrr} \left\lll\int {\rm d^4}x \T\> [{F}_{\mu \nu}(x)
  {F}_{\mu \nu}(x)] \cdot \right.
$$
\be
\left.
\T\> \P [i\int_C {A}_\mu d x_\mu]\right\rrr
-
\left\lll\int {\rm d^4}x \T\> [{F}_{\mu \nu}(x){F}_{\mu
  \nu}(x)]\right\rrr \;
  \label{sa2}
   \ee
 If the loop $C$ is a rectangle with
spatial extension $-R/2<x_3<R/2$ and temporal extension $-T/2
< x_4 < T/2 $ the right-hand side of (\ref{sa2}) can be simplified
further in the limit of large $T$. Let the length $T_g$ be of the order
of the correlation length of the gluon field strengths. Then for
$|x_4|> T/2 + T_g$ the r.h.s. of (\ref{sa2}) is zero and for $|x_4| < T/2 -T_g$
independent of $x_4$. This gives in the limit of large $T$:

\be 8\pi \al_{s0}^2 \frac{\partial\log[\, \lll W[C]\rrr\,
  ]}{\partial \al_{s0}} =  \left\lll
\int {\rm d^3}x \T [{F}_{\mu \nu}(x) {F}_{\mu
  \nu}(x)]\right\rrr_R \> \cdot T  \;
  \label{sa3}
  \ee

 Here and in the following $x_4$ is zero and
the expectation value $\lll .. \rrr_R$ shall denote the expectation
value in the presence of a static quark-antiquark pair at distance
$R$, with the static sources in the fundamental representation, and the
expectation value in absence of the sources has been subtracted.

The potential $V(R)$ is defined by (\ref{eqv1})
and thus one obtains the low
energy theorem:
\be
  -4\pi\al_{s0}^2 \frac{\partial V(R)}{\partial \al_{s0}} =
\frac{1}{2}\left\lll \int {\rm d^3}x\T [{F}_{\mu \nu}(x){F}_{\mu \nu}(x)
  ]\right\rrr_R \;
  \label{7}
  \ee
In (\ref{7}) self energies are understood to be subtracted.

Up to now we have used the bare coupling $g_0$ and bare fields. If we
use the background gauge fixing \cite{Wit671,Wit672,Wit673} we have

\be
  g=Z_g^{-1}g_0\; ,
\; A_\mu = Z^{-1/2}A _\mu^{(0)}\; , \; Z_g Z^{1/2}=1\; , \;
{A}_\mu = g_0 A^{(0)}_\mu = g A_\mu
  \label{gauge}
\ee
where $g$
and $A_\mu$ are renormalized quantities. The renormalized squared
field strength tensor is given by:

\be
  \T [F_{\mu \nu}(x)F_{\mu
  \nu}(x)]=Z_{F^2}^{-1}\T [F^{(0)}_{\mu \nu}(x)F^{(0)}_{\mu
  \nu}(x)]\;
  \label{gauge4}
\ee
where $Z_{F^2}$ was calculated \cite{GriR89} in
background gauge with Landau gauge fixing to be
$  Z_{F^2}=1+\frac{1}{2}g_0\frac{\partial}{\partial g_0} \log Z $
and thus we obtain
  \be \frac{\partial \al_s}{\partial \al_{s0}}= Z Z_{F^2}\; .
\label{gauge3}
  \ee
We use (\ref{gauge})-(\ref{gauge3}) to express (\ref{7}) in terms
of renormalized quantities:

\be -\al_{s}\frac{\partial V(R)}{\partial \al_{s}}
= \frac{1}{2}\left\lll \int {\rm d^3}x\T [F_{\mu \nu}(x) F_{\mu\nu}(x)]\right\rrr_R
\label{7a}
  \ee
In order to evaluate the left-hand side of (\ref{7a})
we use standard renormalization group arguments: the only scales are
the renormalization scale $\mu$ and $R$, thus the potential $V=V(R,\mu
,\al_S)$ satisfies on dimensional grounds:
\be \left( R\frac{\partial}{\partial
  R}- \mu\frac{\partial}{\partial \mu}\right)V(R,\mu ,\al_s)=
  - V(R,\mu ,\al_s)
\label{gauge5}
\ee
But for $V$ as a physical quantity the total
derivative with respect to $\mu$ must vanish:

\be \mu\frac{\rm d}{\rm
  d\mu} V(R, \mu, \al_{s}) = \left( \mu\frac{\partial}{\partial \mu}+
\beta(\al_s)\frac{\partial}{\partial \al_s}\right)
  V(R,\mu ,\al_s)= 0
\label{8}
  \ee
 where $\beta(\al_{s}) \equiv \mu \frac{\rm d}{\rm d\mu}
\al_{s}(\mu)$. We get from (\ref{gauge5}) and (\ref{8}):

\be \frac{\partial
  V(R)}{\partial\al_{s}} = -\frac{1}{\beta} \left\{ V(R) + R
\frac{\partial V(R)}{\partial R} \right\}
  \label{9}
  \ee
 and finally one obtains with
(\ref{7a}) the low energy theorem
\be \left\{ V(R) + R \frac{\partial
  V(R)}{\partial R} \right\} = \frac{1}{2} \frac{\beta}{\al_{s}} \lll
\int {\rm d^3}x \T [F_{\mu \nu}(x) F_{\mu\nu}(x)]\rrr_R
  \label{10}
  \ee
 or
equivalently

\be \left\{ V(R) + R \frac{\partial V(R)}{\partial R}
\right\}=\frac{1}{2}\frac{\beta}{\al_{s}}\lll \int {\rm
  d^3}x\left(\vec{E}(x)^2+\vec{B}(x)^2 \right)\rrr_R
  \label{11}
  \ee
 The
(\ref{11}) is for the case of a linear potential just one of the
many low energy theorems derived in \cite{ope}. But there
renormalization was only discussed to leading order.

The relation between the potential $V(R)$ and the
energy density $T_{00}(x)$ is 
\be V(R) = \int {\rm d^3} x
\lll T_{00}(x)\rrr_R = \frac{1}{2}\lll \int {\rm d^3} x \left( -\vec E(x)^2 +
\vec B(x)^2 \right)\rrr_R
\label{12}
  \ee
 Note that we are in a Euclidean
field theory, hence the minus sign of the electric field.
Moreover on
the r.h.s. of (\ref{11}) and (\ref{12}) we have renormalized composite
operators.

In the  model under consideration where the 
CM and the transversal component of the
electric field vanishes the potential must be proportional to the spatial
integral over $E_{\parallel}^2$.
In view of the
approximations made concerning path ordering as well as the
arbitrarines of the choice of the reference point this is a very
valuable consistency test (see details in \cite{DNR95}).

In lattice QCD (\ref{11}) and (\ref{12}) are known as Michael's sum-rules
\cite{Mic87}. But in the derivation of the action sum-rule (\ref{11})
scaling of $V(R)$ with $R$ was not taken properly into account and
hence the second term on the l.h.s. of (\ref{11}) is missing in
  \cite{Mic87}.

First consider the region where the potential is linear. Then
(\ref{11}) reads:

\be \frac{1}{2}\frac{\beta}{\al_{s}}\lll \int {\rm
  d^3}x\left(\vec{E}(x)^2+\vec{B}(x)^2 \right)\rrr_R = 2\;V(R)
\label{31}
  \ee
 Since in the model of the stochastic vacuum the static
sources do not modify the CM field, i.e.~$\lll \vec B^2\rrr_R =
0$ , we have consistency between (\ref{12}) and (\ref{31}) only if
\be
{\beta}/\al_s = -2
  \label{32}
  \ee
 From the energy sum-rule (\ref{12})
we have already obtained $\al_s=0.57$, thus consistency of the MSV
requires that (\ref{32}) should be satisfied for this value of $\al_s$.

The $\beta$-function can be expanded in a power series in
$\al_s$:

\be \beta(\al_s) = -\frac{\beta_0}{2 \pi}\al_s^2
-\frac{\beta_1}{4 \pi^2}\al_s^3 -\frac{\beta_2}{64 \pi^3} \al_s^4+ ...
\label{33}
  \ee

The n-loop expansion gives the series up to $\al_s^{\rm
  n+1}$. The first two terms of the power series are scheme
independent, the numerical value of the term proportional to $\al_s^4$
has been evaluated \cite{TVZ80} in $\overline{MS}$. For pure gauge SU(3)
theory we have \cite{RPP94}:
\be
  \beta_0=11, \quad \beta_1 = 51, \quad
{\beta_2}_{\overline{MS}} = 2857
  \label{34}
  \ee

We see that the condition (\ref{32}) yields
$\al_s=1.14$ using $\beta$ on the one loop level, $\al_s=0.74$
on the two loop level and $\al_s=0.64$ on the three loop level in
$\overline{MS}$. This value is already very close to the value of
$\al_s=0.57$ determined from the flux tube calculation using
  (\ref{12}).
If we use for $\beta$ the series (\ref{33}) truncated at order
$\al_s^4$ and impose condition (\ref{32}) for $\al_s=0.57$ and
determine the coefficient $\beta_2$ accordingly we find
  $\beta_2=6250$
i.e. about twice the $\overline{MS}$
value.

It is also easy to see from the above that for distances $R/T_g \leq 1$,
where $T_g
\approx 0.3$ fm the MSV
  does no longer yield a linearly rising potential and hence we have 
$$ \left[ V(R)
+ R\frac{\partial V(R)}{\partial R}\right] \neq 2 V(R)\; , \qquad R/T_g \leq
1 $$
Therefore the conditions (\ref{12}) and (\ref{31}) cannot be fulfilled
simultaneously for $\lll \vec B^2\rrr_R =0$. This does by no means speak
against the model, since in the derivation of the results 
it was stressed that the spatial extension of the loop had to be large
as compared to the correlation length of the field strengths in order
to justify the evaluation of $\vec E^2$ and $\vec B^2$.

Notice, that no use of the equations
of motion has been made explicitly. The factorization hypothesis
(Gaussian measure) enters crucially in the evaluation of the squared
color fields. In view of the drastic approximations made it is
gratifying that relation (\ref{32}), which is a highly non-trivial
consequence of the dynamics of the system, is so well fulfilled.

The scale $\mu$ at which the MSV works was found to be the one where
$\al_s(\mu )=0.57$  by using the energy sum-rule. With the
help of the action sum-rule we now checked the consistency of the MSV
at this scale and obtained an effective $\beta$-function.

As can be seen from (\ref{11}) and (\ref{12}) the squared field strengths
$\vec{B}^2$ and $\vec{E}^2$ depend on the renormalization scale $\mu$
and hence on $\al_s(\mu)$. We may use relations (\ref{12}) and
(\ref{31}) in order to predict the ratio of
$$
  Q\equiv \frac{\int {\rm d^3}x \lll \vec B(x)^2\rrr_R}{\int {\rm d^3}
  x\lll \vec
E(x)^2\rrr_R}= \frac{2+{\beta}/\al_s}{2-{\beta}/\al_s}
  $$
as a function of the strong coupling $\al_s$.
  This result can be checked in lattice gauge calculations.

\newpage

\section{Heavy quark dynamics and field correlators}

\setcounter{equation}{0} \def\theequation{4.\arabic{equation}}

In this chapter we apply the MFC to heavy quarkonia having in
mind first of all bottomonium and charmonium and, qualitatively,
$s\bar s$ states. For these systems one should take into account both
perturbative and nonperturbative interactions, since the size of
ground states is  around 0.2 and 0.4 Fm  for bottomonium and
charmonium respectively, and the color Coulomb interaction is
dominant below 0.3 Fm. As we shall see in this chapter the MFC in
general and  the Gaussian stochastic approximation in particular are
very well adapted to the  description  of heavy quarkonia.

Actually MFC is here the only analytic method which
  takes into account both
perturbative and nonperturbative interaction
at all distances. The first account for the 
effects of finite correlation length $T_g$ was made in \cite{lat12}
in attempt to correct the unphysical $n^6$-growth of 
NP energy shift obtained in \cite{6.1,6.2}. 
For states of large
temporal extent (here belong all states of $c\bar c$ and $s\bar s$)
the potential picture emerges from MFC with potentials expressed
through the only two correlators $D$ and $D_1$.
For states of small size the NP configurations can be  accounted for
in form of  condensates, (see Voloshin \cite{6.1} and Leutwyler
\cite{6.2}).
The general situation was discussed in \cite{hq1,hq5}.

Various comparisons and cross-checks can be made with the MFC
results. First, one compares with numerous experimental data and
finds in most cases good agreement; second,  lattice MC
data for  scalar and spin-dependent potentials are available; third,
phenomenological potential models in conjunction with the
${O}(\alpha^2_s)$ perturbative results  are checked versus MFC
predictions.  The reader is also referred to the review \cite{nora2}
where different aspects of the heavy quark dynamics
in the confining vacuum  are extensively
discussed.

The material of present chapter is based mostly on papers
\cite{hq1,hq4} and lectures \cite{shladm} and the paper
\cite{6.7}.

The structure of the chapter is as follows. In section 4.1 the
general formalism of the  Feynman--Schwinger representation (FSR) for
the quark--antiquark  Green's function  is presented and the
nonrelativistic approximation is deduced. In section 4.2 the
NP and perturbative parts of static potential are
derived from the MVC.
In section 4.3 the spin-dependent potentials are obtained from MVC
and compared to the lattice data. Finally, in section 4.4 results of
numerical calculations and general discussion are given.

\subsection{The general formalism for the quark--antiquark
 Green's function}

 We start with the quark Green's function in the form of the proper--time
 and path integral (the Feynman--Schwinger representation
 \cite{hq1,hq2,hq3,hq4,hq5,6.7})
  \be S(x,y)= i(m-\hat{D})\int^{\infty}_0 ds \int Dz\>
  e^{-K}\Upsilon(x,y) \label{6.1}
   \ee
    where $\Upsilon$ contains spin
 insertions into parallel transporter \be \Upsilon (x,y) =
 {\mbox P}_A \> {\mbox P}_F\>
   \exp [i \int^x_y A_{\mu} dz_{\mu}+\int^s_0 d\tau\> g\Sigma
 F(z(\tau))], \label{6.2} \ee $$
 K=m^2s+\frac{1}{4}\int^s_0\dot{z}^2_{\mu}d\tau,
 $$
 and double ordering in $A_{\mu}$ and $F_{\mu\nu}$ is implied by operators
 $P_A,P_F$.

 We have  also
 introduced the $4\times 4$ matrix in Dirac indices
 \be
 \Sigma F \equiv\vec{\sigma}_i
 \left (
 \begin{array}{cc}
 \vec{B}_i& \vec{E}_i\\
 \vec{E}_i&\vec{B}_i
 \end{array}
 \right )
 \ee

Neglecting spins one has instead of (\ref{6.2})
\be
\Upsilon(x,y) \to \Phi(x,y) \equiv \P ~i \int^x_y A_{\mu} dz_{\mu}
\ee
In terms of $(q\bar{q})$ Green's functions (\ref{6.1}) and initial and
final state matrices $\Gamma_i, \Gamma_f$ ( such that
$\bar{q}(x)\Gamma_f\Phi(x,\bar{x}) q(\bar{x}) $ is the final
$q\bar{q}$ state) the total relativistic gauge--invariant $q\bar{q}$
Green's function in the quenched approximation is
 $$
  G(x\bar{x},
y\bar{y})=\lll \T (\Gamma_fS_1(x,y)\Gamma_i\Phi(y,\bar{y})
S_2(\bar{y},\bar{x})\Phi(\bar{x},x))\rrr
$$
\be
-\lll \T(\Gamma_fS_1(x,\bar{x})\Phi(\bar{x},x))
\T (\Gamma_iS_2(\bar{y},y)\Phi(y,\bar{y}))\rrr
 \label{6.5}
 \ee
The angular brackets in (\ref{6.5}) imply averaging over gluon field
$A_{\mu}$.

The second term on the r.h.s. of (\ref{6.5}) represents a mixing of a
flavour singlet quarkonium with two--  or three--gluon glueball. Except
for an accidental degeneration of masses, this term is a small
correction and will be omitted in what follows.

Since we are interested primarily in heavy quarkonia, it is reasonable to do
a systematic nonrelativistic approximation. To this end we introduce
the real evolution parameter (time) $t$ instead of the proper
time $\tau$ in $K,(\bar{\tau}$ in $\bar{K}$) and dynamical mass
parameters $\mu,\bar{\mu}$
 \be
  \frac{dt}{d\tau}= 2\mu_1,\;
\frac{dt}{d\bar{\tau}}=2\mu_2;~~ \int^s_0\dot{z}_{\mu}^2(\tau)d\tau=
2\mu_1\int^T_0dt\left(\frac{dz_{\mu}(t)}{dt}\right)^2
\ee
Here we have denoted
$
T\equiv (x_4+\bar{x}_4)/2
$.
Nonrelativistic approximation means, that one writes for
$z_4(t),\bar{z}_4(t)$
\be
z_4(t)=t+\zeta (t),~~2\mu_1=\frac{T}{s_1};
\label{6.8}
\ee
$$
\bar{z}_4(t)=t+\bar{\zeta}(t),~~2\mu_2=\frac{T}{s_2}
$$
and expands in fluctuations $\zeta, \bar{\zeta}$, which are
${\cal O}(\frac{1}{\sqrt{m}})$ . Note that
the integration in $ds_1 ds_2$ goes over
into $d{\mu}_1d\mu_2$.
Physically
expansion (\ref{6.8}) means that we neglect
in lowest approximation  trajectories with
backtracking of $z_4, \bar{z}_4$, i.e. neglect $q\bar{q}$ pair creation. One
can persuade oneself that insertion of
(\ref{6.8}) into $K,\bar{K}$ allows to
determine $\mu_1,\mu_2$ from the extremum in $K,\bar{K}$ to be
\be
\mu_1=m_1+ {\cal O}(1/m_1),
~~\mu_2=m_2 + {\cal O}(1/m_2)
 \ee
 and one can further make a systematic expansion in powers $1/m_i$. (For
 details see \cite{hq1}). At least to lowest orders in $1/m_i$ this
 procedure is equivalent to the standard (gauge--noninvariant)
 nonrelativistic expansion.

 Let us keep the leading term     of this expansion
 \be
 G(x\bar{x}, y\bar{y})= 4m_1m_2e^{-(m_1+m_2)T}\int D^3zD^3\bar{z}
 e^{-K_1-K_2} \lll W(C)\rrr
 \label{6.10}
 \ee
 where $K_1=\frac{m_1}{2}\int^T_0\dot{z}_i^2(t)dt,~~
 K_2=\frac{m_2}{2}\int^T_0\dot{\bar{z}}_i^2(t)dt$
 and $ \lll W(C)\rrr $ is the W--loop
 operator with closed contour $C$ comprising
 $q$ and $\bar{q}$ paths, and initial and final state parallel transporters
 $\Phi(x,\bar{x})$ and $\Phi(y,\bar{y})$.

 The representation (\ref{6.10})
 is our main object of study in  next
 section.

\subsection{Perturbative and nonperturbative static potentials}

 The technical tool we are using is the background perturbation theory.
 We represent the total gluon field $A_{\mu}$ as
 \be
 A_{\mu}=B_{\mu}+a_{\mu}
 \label{6.11}
 \ee
 where $B_{\mu}$ is the NP background, while $a_{\mu}$ is perturbative
 fluctuation. The principle of  separation in
  (\ref{6.11}) is immaterial for our
 purposes
    (for more discussion of
 this point see \cite{6.6}).

  For static charges trajectories are fixed straight lines and
  $\lll W(C)\rrr $ in (\ref{6.10}) factorizes  out defining the potential
  between static charges $V(R)$
  \be \lll W(C)\rrr = \exp (-V(R)T)
 \label{6.12}
 \ee
   where $R$
  is the distance between charges and $T$ is the time distance in the
  rectangular loop; $C=R\times T$.

  In the approximation, when only bilocal correlators are  kept,
  $V(R)$ is
  given as \cite{hq4,hq5}
  $$
  V_0(R)=C_D \left(-\frac{\alpha_s(R)}{R}+
  2R\int^R_0 d\lambda\int^{\infty}_0 d\nu D(\lambda,\nu)+\right.
  $$
  \be
  \left.+\int^R_0\lambda d\lambda\int^{\infty}_0 d\nu (-2D(\lambda,\nu)
  +D_1(\lambda,\nu))\right)
 \label{6.16}
  \ee

In addition one also obtains radiative corrections due to transverse gluon
exchange, which have been computed earlier \cite{6.10}.

 Thus one has the resulting potential
 \be
 V(R)=
 V_0(R)+\Delta V(R)
 \ee
 where $V_0(R)$ is defined in (\ref{6.16}), and $\Delta V(R)$ is the
 sum of higher--order terms, namely
 $$\Delta V(R)= \Delta V_1(R) +\Delta V_2(R) +\Delta V_3 (R),
 $$
 where $\Delta V_1(R)$ refers to ${\cal O}(\alpha^2_s)$
 perturbative  contributions  \cite{6.111,6.112}, $\Delta V_2(R) $ is
 interference term \cite{6.12} which  is ${\cal O}(\alpha_s)$ and $\Delta
 V_3(R)$ refers to contribution of higher FC.

\subsection{Spin--dependent potentials}

As was shown in the previous section,
  the leading term in the heavy mass
expansion is
the nonrelativistic  form
containing the W--loop with the contour $C$, formed by
straight--line paths without spin insertions ($\Sigma F)$, Eq.
(\ref{6.10}).  To keep spin dependent terms we shall use (\ref{6.2}), (\ref{6.5})
and derive from it the
nonrelativistic expression in
 the limit when $m_i\to \infty$ and also $|\vec x-\vec
y|\ll  |x_0-y_0|$.

Now one can  use the standard  definitions  of
Eichten, Feinberg and Gromes (EFG)
 \cite{6.131,6.132} for the SD
potentials
 $$ V_{SD}(R)=\left(\frac{{\bf \sigma}_1{\bf  L}_1}{4m_1^2}-
\frac{{\bf \sigma}_2{\bf
L}_2}{4m_2^2}\right)\left(\frac{1}{R}\frac{d\varepsilon}{dR}+
\frac{2dV_1(R)}{RdR}\right)+
$$
\be
+
\frac{{\bf \sigma}_2\vec
L_1-
{\bf \sigma}_1\vec
L_2}{2m_1m_2}\frac{1}{R}\frac{dV_2(R)}{dR}+\frac{
{\bf \sigma}_1
{\bf \sigma}_2V_4(R)}{12m_1m_2}+\frac{(3
{\bf \sigma}_1
{\bf R}{\bf \sigma}_2
{\bf R}-{\bf\sigma}_1
{\bf \sigma}_2 R^2)V_3(R)}
{12m_1m_2 R^2}
\label{6.36}
\ee
and identify  $V_i(R)$ by expanding the exponents of
$(\Sigma_i F)$ in (\ref{6.2}) to the first order and  keeping $m^{-2}$
terms as compared to the leading spin--independent (static)
contribution.

In Gaussian (bilocal) approximation one
obtains (we denote quantities taken in Gaussian 
approximation by primes)
   \be
\frac{1}{R}\frac{d{V_1}'}{dR}=-\int^{\infty}_{-\infty}d\nu\int^R_0
\frac{d\lambda}{R}
\left(1-\frac{\lambda}{R}\right)D(\lambda,\nu),
\label{6.41}
\ee

\be
\frac{1}{R}\frac{d{V_2}'}{dR}=\int^{\infty}_{-\infty}d\nu\int^R_0
\frac{\lambda d\lambda}{R^2}
\left[D(\lambda,\nu)+D_1(\lambda,\nu)+\lambda^2\frac{\partial
D_1}{\partial\lambda^2}\right],
\label{6.42}
\ee

\be
\frac{1}{R}\frac{d{\varepsilon}'}{dR}=
\frac{1}{R}
\int^{\infty}_{-\infty} d\nu
\left( \frac{R}{2}D_1(R,\nu)+\int^R_0 d\lambda
D(\lambda,\nu)\right),
\label{6.43}
\ee

\be
V_3=-\int^{\infty}_{-\infty} d\nu R^2\frac{\partial
D_1(R,\nu)}{\partial R^2},
\label{6.44}
\ee
\be
V_4=\int^{\infty}_{-\infty}d\nu
\left(3D(R,\nu)+3D_1(R,\nu)+2R^2\frac{\partial D_1}{\partial R^2}\right)
\label{6.45}
\ee

It is very simple to check, as it was done in \cite{hq3}, that the Gromes
relation
\be
 V'_1-V'_2+\varepsilon ' =0
  \label{6.46}
   \ee
 is identically satisfied in bilocal approximations
for $\varepsilon', V'_1,V'_2$  given by (\ref{6.41}), (\ref{6.42}) and
(\ref{6.43}) for the arbitrary choice of  $D$ and $D_1$ falling off
at large distances.

  One can also check that (\ref{6.43}) is the derivative of the NP part  of
  the scalar potential (\ref{6.16}) (without Coulomb part).

 It is interesting that (\ref{6.41}) and (\ref{6.42})
 yield correct results in lowest order of perturbation theory, namely $V'_1$
 has no contribution and $\frac{V'_2}{R}$  is equal to $\frac{4\alpha_s}{3
 R^3}$, when the perturbative value of $D_1$,
~$D_1^{pert}(x)=16\alpha_s/(3\pi x^4)$, is used.

\subsection{Charmonium and bottomonium spectra.}

In previous sections the scalar and spin--dependent potentials  have been
expressed through the Gaussian FC,  functions $D$  and $D_1$. 
The use of these potentials in the Schr\"odinger equation allows to
calculate the spectra
 of charmonium and bottomonium    and compare results with
experimental data.
The contents of this section is mostly based on \cite{6.15}.

The static potential according to can be written as
\be
V_0(r)=V^{pert}(r)+\varepsilon (r)
\ee
The NP component of $D, D_1$ are taken from
lattice measurements \cite{lat2} in the form (see Section 2)
\be
D(x)=D(0)e^{-|x|/T_g},~~
D_1(x)=D_1(0)e^{-|x|/T_g}
\label{6.23}
\ee
where $D(0) \approx 3D_1(0)$.

Parameters  used
above for computations and corresponding to
lattice results \cite{lat12,lat2} are

 $$
\sigma =0.185\> {\mbox{GeV}}^2,~
~\sigma_1 = \frac{1}{\pi} D_1(0)T_g^2 = 0.0064\> {\mbox{GeV}}^2,
{\mbox{GeV}}   $$
\be
T_g ^{-1} = 1\>\mbox{GeV}, ~~ 
m_b=4.85\> \mbox{GeV},~~ m_c=1.45\> \mbox{GeV} 
\label{6.24}
\ee

The perturbative part of potential takes into account the asymptotic freedom
and freezing at large distances
with $\alpha_s=\alpha_s^{max}=0.5$ \cite{frez1,frez2,frez3}.

Results of computations for  the parameters of Eq. (\ref{6.24})
 are listed below in the Tables 1-2 together with experimental
 data and calculations with the Cornell potential \cite{6.14}.\\
  \newpage

{\bf Table 1}\\

\begin{center}

Masses of $b\bar b$ for
FC (\ref{6.23}) in comparison with experiment and
those of Cornell potential \cite{6.14}\\
\bigskip

\begin{tabular}{|l|r|r|r|}\hline
State& Exper.&  FC& Cornell\\ \hline
$\gamma (1S)$ & 9460 &  9460 & " fit"\\
$\gamma (2S)$ & 10023 &10023 &10052\\
$\gamma (3S)$ & 10355 &10347 &10400\\
$\chi_b(1P)$ &  9900 &9918 &9960\\
$\chi_b(2P)$ & 10260 &10250& 10314\\
$\gamma(2S)-\gamma(1S)$ & 563 &563  &592   \\
$\gamma(3S)-\gamma(2S)$ &332    &324  &345   \\
$\gamma(2S)-\chi_b(1P)$ &123    &105  & 93   \\
$\chi_b(2P)-\chi_b(1P)$ &359    &332  &354   \\\hline
 \end{tabular}

~~~~\\

\end{center}


{\bf Table 2}\\

\begin{center}

Masses of $c\bar c$ for FC (\ref{6.23}) $vs$ experiment \\
\bigskip

\begin{tabular}{|l|r|r|r|}\hline
State& Exper.& FC& Cornell\\ \hline
$J\psi$ & 3097 &  3096& 3097\\
$\Psi(2S)$ & 3686 &  3682&3688\\
$\Psi(3S)$ & 4040 & 4128& 4111\\
$\chi_{cog}(1P)$ & 3525 &  3516& 3525\\
$\Psi(1D)$ & 3770 &  3800& 3810\\
$\Psi(2S)-J/\psi'$ & 589  &  586& 591 \\
$\Psi(2S)-\chi_{cog}(1P)$ & 161  &  166&163 \\
$\Psi(1D)-J/\psi'$ & 673  &  705 & 713\\ \hline
 \end{tabular}

\end{center}

Results listed in Tables 1-2  demonstrate several successes of MFC
in comparison with  
the standard potential models like Cornell.
Namely, {\bf i)} it is possible to describe bottomonium spectrum with the current
$b$-quark mass, $m_b=4.8 \> GeV$, while in Cornell potential an effective mass
$m_b=5.17 \> GeV$ is used; {\bf ii)} a good description  of leptonic width is 
obtained (not shown in Tables 1,2, see \cite{6.15});
{\bf iii)} the overal agreement of
levels with experiment is rather good and certainly better than for
the Cornell potential.

The fine structure of the spectrum is much more delicate
issue. In particular, here one should use the ${\cal O}(\alpha^2_s)$ contributions to
LS and scalar interaction.
 The full analysis done in \cite{6.18} demonstrates a good description of charmonium
and also checks the
consistency of the model 
with the lattice data on spin-dependent potentials \cite{6.17}.

\newpage

\section{Field correlators at $T > 0$ and deconfinement transition
}

\setcounter{equation}{0} \def\theequation{5.\arabic{equation}}

The finite temperature QCD provides
a unique insight into the structure
        of the QCD vacuum, in particular it
gives an important information about
the mechanism of
confinement. It is also of more practical
interest, since the nature of the  deconfinement transition has
its bearing on the cosmology  and the hot
QCD plasma can possibly  be tested in heavy ion collisions.

 From  theoretical point of view the  hot
QCD is a unique theoretical laboratory where both perturbative (P)
 and  nonperturbative (NP) methods can be
applied in different temperature regimes.

It is believed that the perturbative QCD is applicable in the deconfined
phase at large enough temperatures $T$, where the effective coupling
 constant $g(T)$ is small \cite{7.11,7.12}, while at small $T$ (in the confined
 phase) the NP effects instead are most important. However even at
 large T  the physics is not that simple: some effects, like
 screening (electric gluon mass), need a resummation of the
 perturbative series \cite{7.21,7.22}, while the effects connected with the
 magnetic gluon mass demonstrate the infrared divergence of the
 series \cite{7.3}. For the role of hadron degrees of freedom in the
transition see e.g. \cite{7.19}.

During last years there appeared a lot of lattice data which point
at the NP character of dynamics above $T_c$. Here
belong: {\bf i)} area law of spatial W--loops \cite{7.41,7.42},  {\bf ii)} screening masses
of mesons and baryons \cite{7.51,7.52} and
glueballs \cite{7.6},  {\bf iii)} temperature
dependence of Polyakov--line correlators \cite{7.7}.

In addition, behaviour of $\epsilon-3p$ above $T_c$ has a bump
incompatible with the simple quark--gluon  gas picture
\cite{7.8}.

Thus the inclusion of NP configurations into
analysis at $T > 0$ and also at $T > T_c$ is necessary.
This was done using background field method in \cite{7.9}. Besides
defining NP dynamics, the NP configurations also ensure freezing of
$\alpha_s$ and stabilize perturbative series \cite{7.9}. To describe the
phase transition a simple choice of deconfined phase was suggested
\cite{7.10} where all $NP$ CM configurations are kept intact
as in the confined phase, whereas CE correlators
responsible for confinement, vanish.

 Recently this picture of phase transition has been tested and
confirmed in the lattice calculations  \cite{lat31,lat32} of electric and
magnetic correlators. It was shown in
\cite{lat41,lat42,lat43} that the confining part of the electric correlator $D^E$
decreases steeply right above $T_c$, while the deconfining
electric part, $D^E_1$, and magnetic part $D^B, D^B_1,$ stay
approximately constant.
see Figs.2,3. Thus one has a consistent mechanism
describing the structure of the QCD vacuum at $T\geq 0 $ and the
phase transition at $T=T_c$.
 This
picture together with the background perturbation theory forms a
basis of quantitative calculations, where field correlators
(condensates) are used as the $NP$ input \cite{7.10}.

The plan  of
the Section is as follows. In subsection 5.1 the modified
background field formalism
\cite{7.9} is presented, based on the familiar
background field method
\cite{Wit671,Wit672,Wit673} for $T > 0$ and
incorporating the 't Hooft's identity for integration over quantum
fluctuations as well as over the  background fields.  The temperature
phase transition is discussed in subsection 5.2 and resulting predictions
for $T_c$  are compared with lattice data.  In subsection 5.3 the spacial
  W--loops are computed both below and above $T_c$.

\subsection{Background field formalism}

 We derive here basic formulas for the partition function, free energy and
Green's  function in the NP background formalism at
$T > 0$ \cite{7.10}. The total gluon field $A_{\mu}$ is split into a
perturbative part $a_{\mu}$ and NP background $B_{\mu}$
\be
A_{\mu}=B_{\mu}+a_{\mu}
\ee
where both $B_{\mu}$ and  $a_{\mu}$ are subject to periodic boundary
conditions. The principle of this separation is immaterial for our purposes
here, and one can average over fields $B_{\mu}$ and $a_{\mu}$ independently
\be Z=\int DA_{\mu}\> \exp (-S(A)) =
\frac{\int DB_{\mu}\eta(B)\int Da_{\mu} \exp (-S(B+a))}{\int DB_{\mu}\eta(B)}
\label{12.2}
\ee
  $$ \equiv\lll\lll 
\exp(-S(B+a)){\rrr}_a{\rrr}_B $$
   with the weight $\eta(B)$. In our case we choose $\eta(B)$ to fix field
correlators and string tension at their observed values.

The partition function can be written as $$ Z(V,T,\mu=0) =
\lll Z(B){\rrr}_B$$
where
\be
Z(B)=N\int Da_{\mu}\> \exp (-\int^{\beta}_0 d\tau \int d^3x\> L(x,\tau))
\label{12.3}
\ee
 where $\phi$ denotes all set of fields $a_{\mu}, \Psi, \Psi^+$; $N$ is a
normalization  constant, and the brackets $\lll ... {\rrr}_B$  means some averaging over
(nonperturbative) background fields $B_{\mu}$ with the weight $\eta(B)$,
 as in (\ref{12.2}). Explicitly 
$$
S(B+a) = \int^{\beta}_0 d\tau \int d^3x\> \sum^{8}_{i=1} 
L_i,$$
where
\begin{eqnarray}
\nonumber
L_1=\frac{1}{4} (F^a_{\mu\nu}(B))^2;\;  L_2=\frac{1}{2} a_{\mu}^a W_{\mu\nu}^{ab} a_{\nu}^b,
\\
L_3=\bar{\Theta}^a (D^2(B))_{ab}\Theta^b;\; L_4=-ig\bar{\Theta}^a (D_{\mu}, a_{\mu})_{ab}\Theta^b
\\
\nonumber
L_5=\frac{1}{2}\alpha (D_{\mu}(B)a_{\mu})^2;\;  L_6=L_{int} (a^3,a^4)
\\
\nonumber
L_7=- a_{\nu} D_{\mu}(B) F_{\mu\nu}(B);\; L_8=\Psi^+(m+\hat{D}(B+a))\Psi
\label{12.4}
\end{eqnarray}
Notice, that we have redefined the fields according to $A_{\mu} \to g A_{\mu}$
in this section.

In (\ref{12.4}) $\bar{\Theta},\Theta$ are ghost fields, $\alpha$ is gauge--fixing constant,
$L_6$
contains 3-- and 4--gluon vertices.
For the fields $B_{\mu}$ satisfying classical equations
of motion $L_7 = 0$.

 The inverse gluon propagator in the background gauge is
\be
W^{ab}_{\mu\nu} =- D^2(B)_{ab} \cdot \delta_{\mu\nu} - 2 g F^c_{\mu\nu}(B) f^{acb}
\ee
where
\be
(D_{\lambda})_{ca} = \partial_{\lambda} \delta_{ca} - ig T^b_{ca} B^b_{\lambda} \equiv
\partial_{\lambda} \delta_{ca} - g f_{bca} B^b_{\lambda}
\ee
We consider first the case of pure  gluodinamics, $L_8\equiv 0$.
Integration over ghost and gluon degrees of freedom in (\ref{12.3}) yields
\begin{eqnarray}
\nonumber
Z(B) =N'(\mbox{Det} W(B))^{-1/2}_{reg}\> [\mbox{Det}
(-D_{\mu}(B) D_{\mu}(B+a))]_{a=\frac{\delta}{\delta J}} \times
\\
\times \left\{ 1+ \sum^{\infty}_{l=1} \frac{S_{int}^l}{l!}
\left(a= \frac{\delta}{\delta J}\right) \right\}
\>\exp \left(-\frac{1}{2} J
W^{-1}J\right)_{J_{\mu}= \;D_{\mu}(B)F_{\mu\nu}(B)}
\label{12.7}
\end{eqnarray}

One can consider  strong background fields, so that $gB_{\mu}$ is large (as
compared to $\Lambda^2_{QCD}$), while $\alpha_s=\frac{g^2}{4\pi}$
in that strong background is small at all distances \cite{frez1,frez2}.
In this case (\ref{12.7}) is a perturbative sum in powers of $g^n$,
arising from expansion in $(ga_{\mu})^n$.
In what follows we shall discuss the Feynman graphs for the free energy $F(T),$
connected to $Z(B)$ via
\begin{eqnarray}
F(T) = -T \>\log \lll Z(B){\rrr}_B
\end{eqnarray}

As will be seen, the lowest  order graphs already  contain  a nontrivial
dynamical mechanism for  the deconfinement transition.
To the lowest order in $ga_{\mu}$ the partition function and free
energy are
$$ Z_0=\lll \exp(-F_0(B)/T){\rrr}_B
$$
$$
F_0(B)/T=\frac{1}{2} \log
\>\mbox{Det} W- \log \> \mbox{Det} (-D^2(B))= $$
\be
= \T \int^{\infty}_0\zeta(t)\frac{dt}{t}\left(-\frac{1}{2}e^{-tW}+
e^{tD^2(B)}\right)
\ee
where $\hat{W}=-D^2(B)-2g\hat{F}$ and $D^2(B)$ is the inverse gluon and
ghost propagator  respectively, $\zeta(t)$ is a regularizing factor
\cite{7.10}.

The ghost propagator can be written as \cite{7.10,7.131,7.132}
\be
(-D^2)^{-1}_{xy}=\lll x|\int^{\infty}_0 ds\> \exp(sD^2(B))|y\rrr =
\int^{\infty}_0ds(Dz)^w_{xy}e^{-K}\Phi(x,y)
\label{12.10}
\ee
where winding path integral is introduced \cite{7.10}
\be
(Dz)^w_{xy}=\lim_{N\to
\infty}\prod^{N}_{m=1}\frac{d^4\zeta(m)}{(4\pi\varepsilon)^2}
\sum^{\infty}_{n=0,\pm1,..}
\int\frac{d^4p}{(2\pi)^4}\> e^{ip\left(\sum\zeta(m)-(x-y)-n\beta
\delta_{\mu 4}\right)}
\label{12.11}
\ee
with $\beta=1/T$. For the gluon propagator an analogous expression holds
true, expect that in (\ref{12.4}) one should insert gluon spin factor
${\mbox{P}}_F \exp\>2g\hat{F}$ inside $\Phi(x,y)$. For a
quark propagator the sum over windings
acquires the factor $(-1)^n$ and quark spin factor is $\exp
\>(g\sigma_{\mu\nu}F_{\mu\nu})$ \cite{7.10}.

\subsection{Temperature phase transition in QCD}

We are now in position to make expansion of $Z$ and $F$ in powers of
$ga_{\mu}$ (i.e. perturbative expansion in $\alpha_s$), and the
leading--nonperturbative term $Z_0, F_0$ -- can be represented as a sum of
contributions with different powers of $N_c$ and we 
systematically will
keep the leading terms ${\cal O}(N_c^2),{\cal O}(N_c)$ and
${\cal O}(N_c^0)$.

To describe the temperature phase transition one should specify
phases and compute free energy. For the confining phase to lowest
order in $\alpha_s$ free energy is given by
(\ref{12.3}) plus contribution
of energy density $\varepsilon $ at zero temperature
\be
F(1)=\varepsilon V_3-\frac{\pi^2}{30}V_3T^4-T\sum_s\frac{V_3(2m_{\pi}
T)^{3/2}}{8\pi^{3/2}}\exp(-\frac{m_{\pi}}{T})+{\cal O}(1/N_c)
\label{12.12}
\ee
where $\varepsilon$ is defined by scale anomaly
\be
\varepsilon \simeq
-\frac{11}{3}N_c\frac{\alpha_s}{32\pi}\lll (F^a_{\mu\nu}(B))^2\rrr
\ee
and the next terms in (\ref{12.12})
correspond to the contribution of mesons (we
keep only pion gas) and glueballs. Note that $\varepsilon={\cal O}(N^2_c)$
while two other terms in (\ref{12.12}) are ${\cal O}(N^0_c)$.

For the second phase (to be the high temperature phase) we had made an
assumption (which was proved on the lattice \cite{lat41,lat42})
 that all CM field correlators are the same as
in the first phase, while all CE fields vanish. Since at
$T=0$ CM correlators  and CE
correlators  are equal due to the Euclidean $O(4)$ invariance,
one has
\be
\lll (F^a_{\mu\nu}(B))^2\rrr =\lll (F^a_{\mu\nu})^2\rrr _{el}+
\lll (F^a_{\mu\nu})^2\rrr _{magn};\;
\lll F^2\rrr _{magn}=\lll F^2\rrr _{el}
\ee

The string tension $\sigma$ which characterizes confinement is due to the
electric fields \cite{ds11,ds12}, e.g. in the plane $(i4)$ one has
\be
\sigma=\sigma_E=\frac{g^2}{2}\int\int
d^2x\lll \T \>E_i(x)\Phi(x,0)E_i(0)\Phi(0,x)\rrr +...
\ee
where dots imply higher order terms in $E_i$ (notice, that we have redefined the fields
$A\to gA$ in this chapter with respect to the notations adopted in 
Section 3.)

Vanishing of $\sigma_E$ liberates gluons and quarks, which will contribute
to the free energy in the deconfined phase by their closed loop terms
(\ref{12.11}) with all possible windings while the
magnetic correlators enter via perimeter
contribution.  As a result one
has for the high-temperature phase (phase 2) (cf.\cite{7.10}).
 \be
F(2)=\frac{1}{2}\varepsilon
V_3-(N^2_c-1)V_3\frac{T^4\pi^2}{45}\Omega_g-\frac{7\pi^2}{180}N_cV_3T^4
n_f\Omega_q+{\cal O}(N_c^0)
\label{12.16}
\ee
where $\Omega_q$ and $\Omega_g$ are perimeter terms for  quarks and
gluons respectively,  the latter was estimated in \cite{7.16} from the
adjoint Polyakov line;
 in what follows we replace
$\Omega$ by one for simplicity.

Comparing (\ref{12.12}) and (\ref{12.16}), $F(1)=F(2)$ at $T=T_c$, one finds in the
order ${\cal O}(N_c)$, disregarding all meson and glueball contributions
\be
T_c^4 =\frac{\frac{11}{3}N_c\frac{\alpha_s\lll F^2\rrr }{32\pi}}
{\frac{2\pi^2}{45}
(N^2_c-1)+\frac{7\pi^2}{90}N_cn_f}
\label{12.17}
\ee
For standard value of $G_2\equiv \frac{\alpha_s}{\pi}\lll F^2\rrr =0.012\>
\mbox{GeV}^4$
 (note that for $n_f=0$ one should use
larger value of $G_2$ \cite{lat0}) one has for $SU(3)$ and
different values of $n_f=0,2,4$ respectively $T_c=~240,$ $150,134\> \mbox{MeV}$.
This should be compared with lattice data \cite{7.8}
$T_c^{lat}=~240,$ $146,131\> \mbox{MeV}$.  Agreement is quite good.  Note that
at large $N_c$ one has $T_c={\cal O}(N_c^{0})$ i.e. the resulting value of
$T_c$ doesn't depend on $N_c$ in this limit. Hadron contributions to
$T_c$ are ${\cal O}(N_c^{-2})$ and therefore suppressed if $T_c $ is below
the Hagedorn  temperature as it typically happens in string theory
estimates  \cite{7.17}.

Till now we disregarded all perturbative and nonperturbative
corrections
to $F(2)$ except for magnetic condensate,
the term $\frac{1}{2}\varepsilon V_3$.
If we disregard also this term,
considering in this way only free
gas of gluons and quarks for the phase 2,
we come to the model, considered in \cite{7.18}.
The values of $T_c$ obtained in this
way differ from ours not much --
they are factor of $2^{1/4}\approx 1.19$ larger, but one  immediately
encounters problems with explanation of spatial string tension,
screening masses etc.,
which are naturally are accounted for by the notion of magnetic
confinement -- nonzero values of magnetic correlators
in the  phase 2, including magnetic  condensate term,
$1/2\varepsilon V_3$.

Notice, that NP corrections may contribute to $\Omega_g,\Omega_q$. Their
phenomenological necessity can be seen in the
measured values of $\varepsilon-3p$, see the corresponding pictures 
in \cite{7.7,7.8}.
In case of $\Omega_g=\Omega_q=1$ the difference  $\varepsilon -
3p$ should be zero, and of course higher orders in $N_c^{-1}(NP$
effects )  and higher orders in $g$ (perturbative effects) contribute
to it.  In \cite{7.16} the authors tried to estimate effect of nonzero
($\Omega_g-1$), which is ${\cal O}(N_c^2)$, on the energy density and
pressure. To this end the adjoint Polyakov line was used and
 the NP perimeter contribution was separated from it.

\subsection{Spacial W--loops}

 In this section we derive  area law for spatial W--loops,
expressing spatial string tension in terms of CMC.

        In standard way one
obtains the area law for large W--loops of  size $L$, $L\gg
T_g^{(m)}$ ($T_g^{(m)}$ is the magnetic correlation  length)
 \be
\lll W(C)\rrr _{spacial}\approx exp (-\sigma_s S_{\min})
\ee
where the  spacial string tension is \cite{7.10}
\be
\sigma_s=\frac{g^2}{2} \int d^2x\lll\lll  B_n(x)\Phi(x,0)
B_n(0)\Phi(0,x)\rrr\rrr + {\cal O}(\lll B^4\rrr )
\label{12.21}
\ee
 and $n$ is the component normal to the plane of the contour,
 while the last term in (\ref{12.21}) denotes contribution  of the fourth
 and higher order cumulants. On general grounds one can write for the
 integrand in (\ref{12.21})(see also section 2):
 $$
 \lll\lll  B_i(x)\Phi(x,0)B_j(0)\Phi(0,x)\rrr\rrr =
 \delta_{ij}\left(D^B(x)+D_1^B(x)+\vec{x}^2\frac{\partial D^B_1}{\partial
 x^2}\right)
$$
\be
- x_ix_j\frac{\partial D^B_1}{\partial x^2},
 \ee
 and only the term $D^B(x)$ enters in (\ref{12.21})
 \be
 \sigma_s=\frac{g^2}{2}\int d^2xD^B(x)+{\cal O}(\lll B^4\rrr )
 \ee
 similarly for the temporal W--loop in the plane $(i4)$ one  has
 the area law for $T < T_c$  with temporal string tension
 \be
 \sigma_E=\frac{g^2}{2} \int d^2xD^E(x)+ {\cal O}(\lll E^4\rrr )
 \ee
 For $T=0$ due to the $O(4)$  invariance CE and CM correlators  
 coincide and
 $\sigma_E=\sigma_s$.
 The lattice measurements of $D_E$ and $D_B$ \cite{lat31,lat32} show that
 $\sigma_E$ and $\sigma_S$ stay approximately constant in the whole
 interval $0\leq T\leq T_c$, as it is suggested by the phase
 transition mechanism described above. For $T > T_c$ in the
 deconfining phase CEC vanish, while CMC stays
nonzero and
 change on the scale of the dilaton mass $\sim 1\> \mbox{GeV}$, therefore one
 expects that $\sigma_s$ stays intact till the onset of the
 dimensional reduction mechanism.  This expectation is confirmed by
 the lattice simulation --- $\sigma_s$ stays constant up to $T\approx
 1.4\> T_c$ \cite{7.20}.  Lattice data \cite{7.20,7.21u} show an
 increase of $\sigma_s$ at $T\approx 2\>T_c$,  for $SU(2)$ which could
 imply the early onset of dimensional reduction.

It was shown how the Feynman--Schwinger representation  (FSR) which
proved to be very useful for $T=0$, is modified for $T > 0$. In
particular, all gluon and quark propagators are written as sums of
path integrals with winding paths in the 4th direction.
Using
FSR the Hamiltonian for screening states of mesons and glueballs is derived,
in \cite{7.22u} and screening masses and wave functions for lowest meson
and glueball states are numerically computed.
All this exemplifies NP
background field theory and in particular MFC  as
a powerful instrument both for $T < T_c$ and for $T > T_c$.

\section{ Conclusions}

 The material of previous chapters contains only a small part  of
 results obtained with the method of FC. The aim of the present
 review is to explain foundations of the method and possible lines of
development.

In all fields  where NP configurations are important the method of FC
can be used to give a systematic description and calculation of
effects. Here is the list of some results obtained heretofore with FC:

\begin{itemize}
\item Proposed dominance of the lowest (Gaussian) FC gives a clue to the
simple stochastic picture of confinement in agreement with lattice
data.
\item  The $q\bar q$ and gluon -- gluon interaction is
 approximately proportional to
 the charge squared (quadratic Casimir operator $C_D^{(2)}$) which is
 supported by lattice data.
\item Heavy quark dynamics (masses, decay rates etc) calculated via FC
 are in good  agreement with experiment.
\item The string profile (field distribution inside the string ) is
 well described by the lowest Gaussian correlator, for both types of
 probes (see chapter 3).
\item The consistency of the model with low-energy theorems is
established.
\item Lattice calculations yield FC with magnitude and correlation
 length $T_g$ which quantitatively correspond to the string tension
 and gluonic condensate known from experiment. The use of lattice
 measurements allows to put all the method of FC on the firm
 foundation, since using FC from lattice data enables one to predict
 hundreds of data: masses, widths, cross sections etc.
\item It was predicted that at the deconfinement phase
 transition the correlator of CE fields $D_E$ vanishes,
 while  that of CM fields $D_B$ stays intact. Lattice
 measurements have confirmed this and have shown the
 dramatic decrease of $D_E$ at $T\ge T_c$, while $D_B$ is almost
 equal to its value at $T=0$. From this fact it follows that $D_E$ is a
 good order parameter for the deconfinement  phase transition, and
 also that above $T_c$ one has the  phase of "magnetic confinement".
\end{itemize}

It is worth noting, that the fundamental input of the formalism --
the Gaussian correlator -- is not a freely adjustable function.
Instead, one gets it from the lattice measurements and 
analytically calculates all the rest without any additional 
"fitting". It is one of the reasons why so much 
attention has been paid throughout this review to the question
about Gaussian dominance, since in this case one basically does
not need any other field-theoretical inputs (like higher correlators).

 One should also mention four topics which might have
 important influence on the development of the method.
An important recent development which has a direct bearing on FC
is the calculation of the so called gluelump states on the lattice 
\cite{micrec}
and in the present method \cite{simrec}.
Actually gluelump Green's functions are generalizations of FC and the
lowest gluelump states, $1^{--}$ and $1^{+-}$ refer directly
to the FC of electric and magnetic fields, respectively.
Therefore the masses of these gluelump states are simply $T_g^{-1}$
for the corresponding correlators.

 The second line of development is the analytic approach with the aim of
construction of the equations,  defining the FC. In this  way one should
 analytically calculate FC
and establish the hierachy of FC corresponding to the Gaussian vacuum.
  Recently a set of integral equations has been derived in the
 large $N_c$ limit \cite{coreq}, and the first calculation of $T_g$ in terms of
 $\sigma$ was made possible, yielding results in good agreement with lattice
 data.  More extended analysis is necessary in this direction including
 phase transition and dynamical quarks.

As mentioned in the introduction, the model has also been successfully
applied to high energy reactions with
small momentum transfer. Here it was possible with the correlator
discussed here to calculate the matrix elements of many high
energy reactions involving hadrons and (virtual) photons. A discussion of
the methods and the results can be found for instance in the reviews 
\cite{revsc1}, \cite{revsc3} and in 
and the literature quoted there. Also this is a wide field and will be
treated separately.

 Finally, there is another area of research important for applications and
 understanding -- the interconnection of FC on one hand and OPE and
 QCD sum rules on another hand, especially from the point of view
of OPE violation.
As it was argued by one of the authors \cite{6.12}
the problem of OPE violation is tightly connected
to a vast and almost unexplored  field of interference
between purely perturbative and purely nonperturbative
contributions. This direction of studies looks promising however
difficult.
The work on this set of questions \cite{frez2,frez3}
  in the example of $e^-e^+$ annihilation
 raises many interesting problems, especially about
ultraviolet renormalons
 and $1/Q^2$ term, which together with recent lattice measurements of
 the condensate $\lll FF\rrr$ (see discussion and refs.
in \cite{march}) makes
 this field highly attractive and dynamic -- hence still
 inappropriate for the review.

\section{Acknowledgements}

H.G.D., V.Sh. and Yu.S. acknowledge the support
from the grant DFG-RFFI-00088G.
A.DiG., V.Sh. and Yu.S. acknowledge the support
from the grants INTAS 01-110.
V.Sh. and Yu.S. acknowledge the support  from the grants   
RFFI-00-02-17836 and RFFI-96-15-96740
for scientific schools. 

 It is a pleasure for the authors to thank their colleagues from 
Institut for Theoretical Physics, Utrecht University; Institut f\"ur Theoretische Physik, Heidelberg
Universit\"at; Humboldt Universit\"at, Berlin; Dipartimento di Fisica,
Pisa University and ITEP theory group for numerous and useful discussions.

\newpage 

\centerline{\bf Figures}

\begin{minipage}{0.9\linewidth}
\epsfxsize0.8\linewidth { \epsfbox{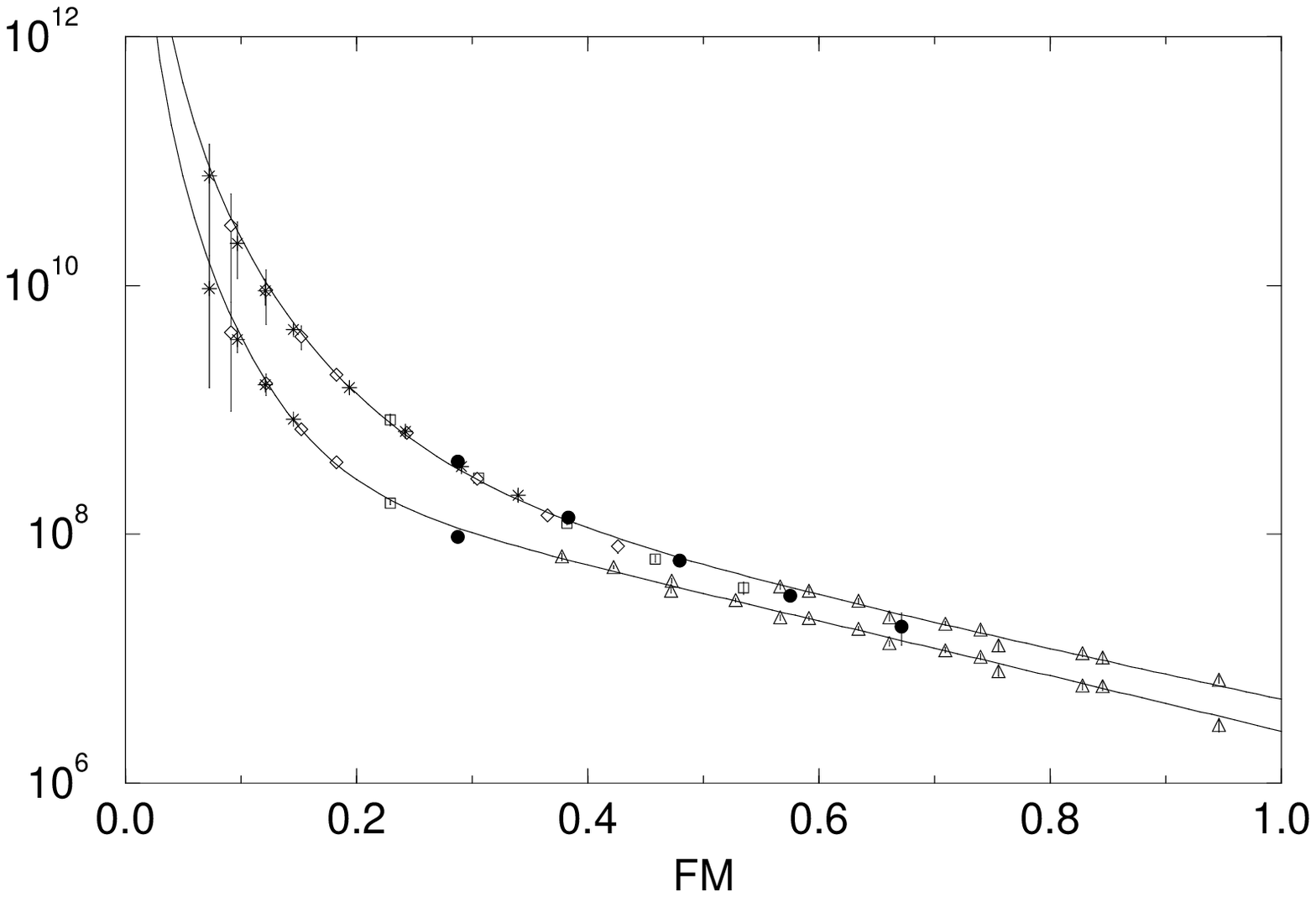}} 
\end{minipage}
\bigskip

\noindent {\bf Fig.1.} $D_{||}^{lat}/a^4 = {D} + {D}_1 + z^2 \partial D_1/\partial
z^2$
and $D_\perp/a^4 = {D}+ {D}_1$ versus $z$. The lines correspond to
the best fit to (\ref{adg2}) (from \cite{lat0}).

\newpage

\begin{minipage}{0.9\linewidth}
\epsfxsize0.8\linewidth
\rotatebox{270}{
\epsfbox{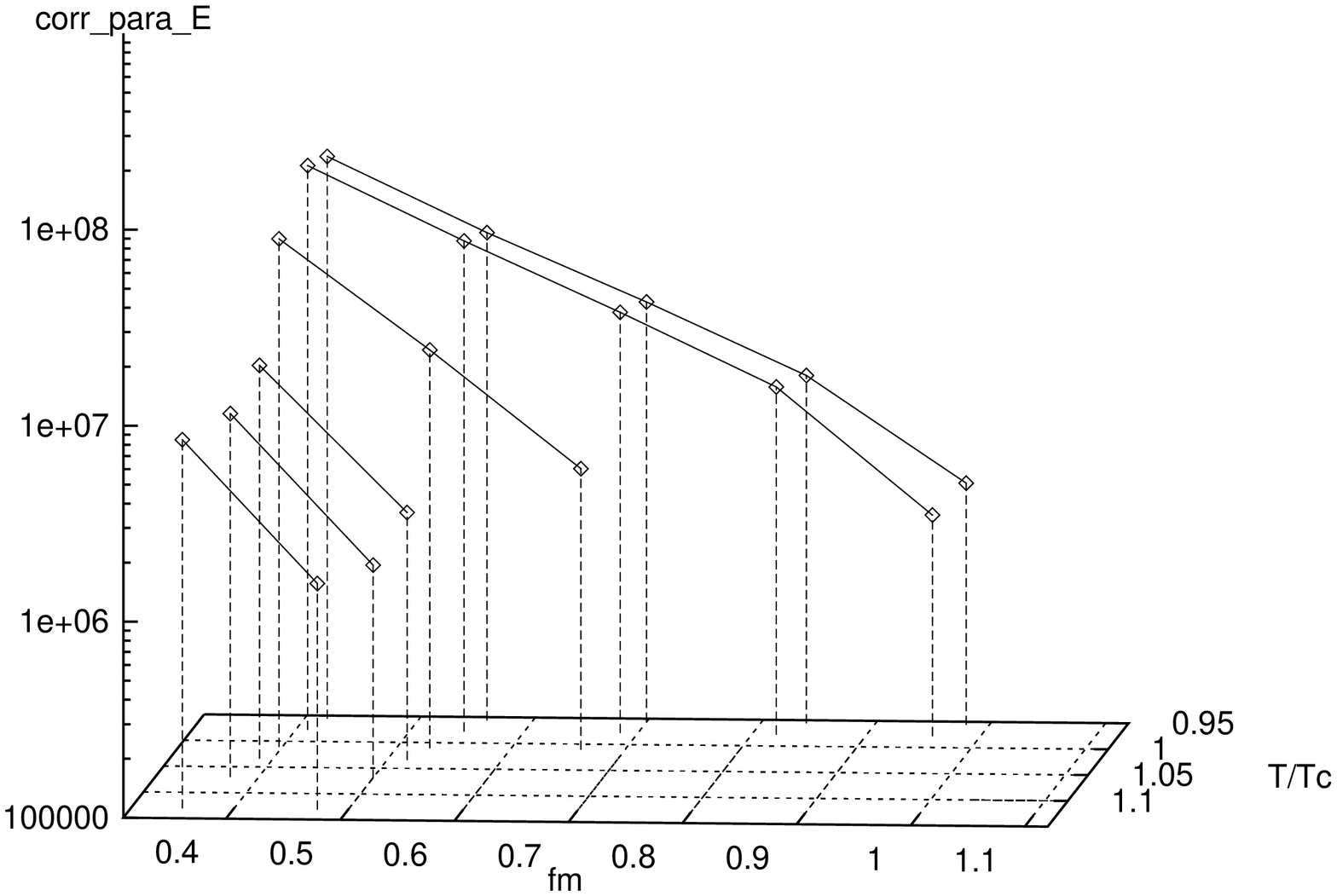}}
\end{minipage}
\bigskip 

\noindent {\bf Fig.2.} The electric longitudinal correlator versus distance,
    for different values of $T/T_c$ (from \cite{lat0}).

\newpage

\begin{minipage}{0.9\linewidth}
\epsfxsize0.8\linewidth
\rotatebox{270}{
\epsfbox{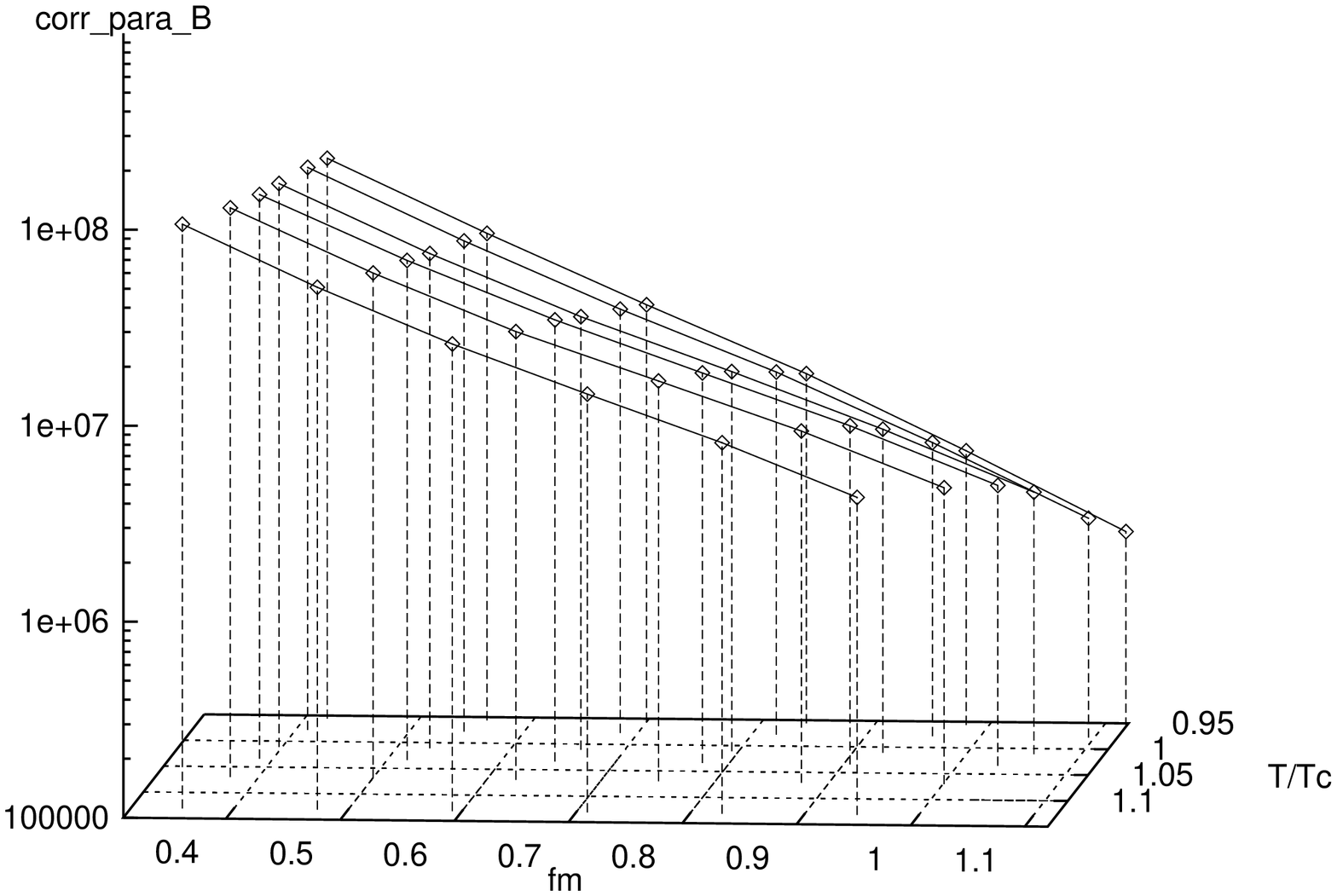}}
\end{minipage}
\bigskip

\noindent {\bf  Fig.3.} The magnetic longitudinal correlator versus distance,
for different values of $T/T_c$ (from \cite{lat0}).

\newpage

\begin{minipage}{0.9\linewidth}
\epsfxsize0.9\linewidth { \epsfbox{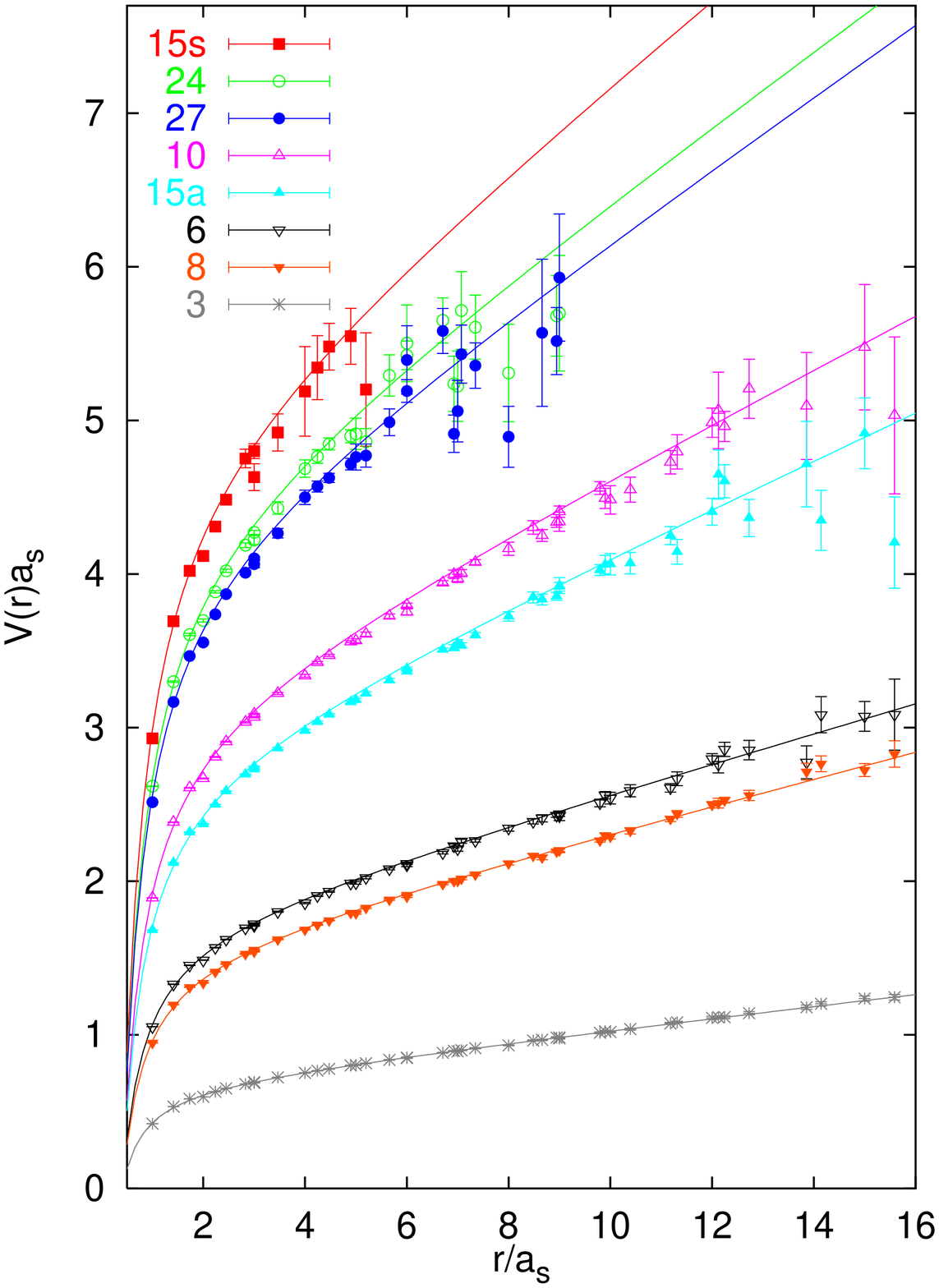}} 
\end{minipage}
\bigskip

\noindent {\bf Fig.4.} Static potentials for sources in different $SU(3)$
representations (from \cite{bali3}).

\newpage

\begin{minipage}{0.9\linewidth}
\epsfxsize0.9\linewidth { \epsfbox{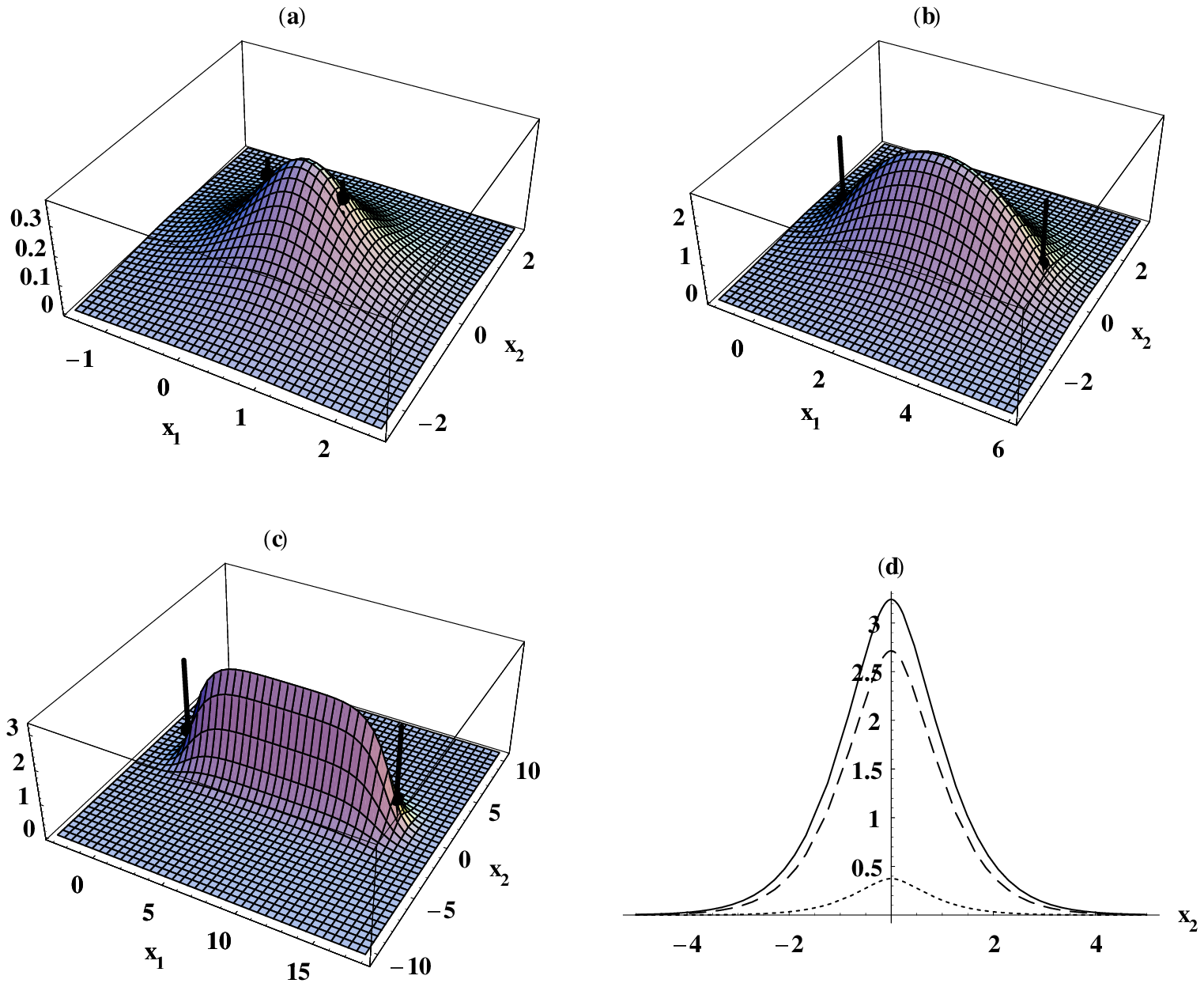}} 
\end{minipage}
\bigskip

\noindent {\bf Fig.5.} String profiles measured with connected correlators 
for different distances (from \cite{kuzm}).

\newpage

\begin{minipage}{0.55\linewidth}
\epsfxsize0.7\linewidth
\epsfbox{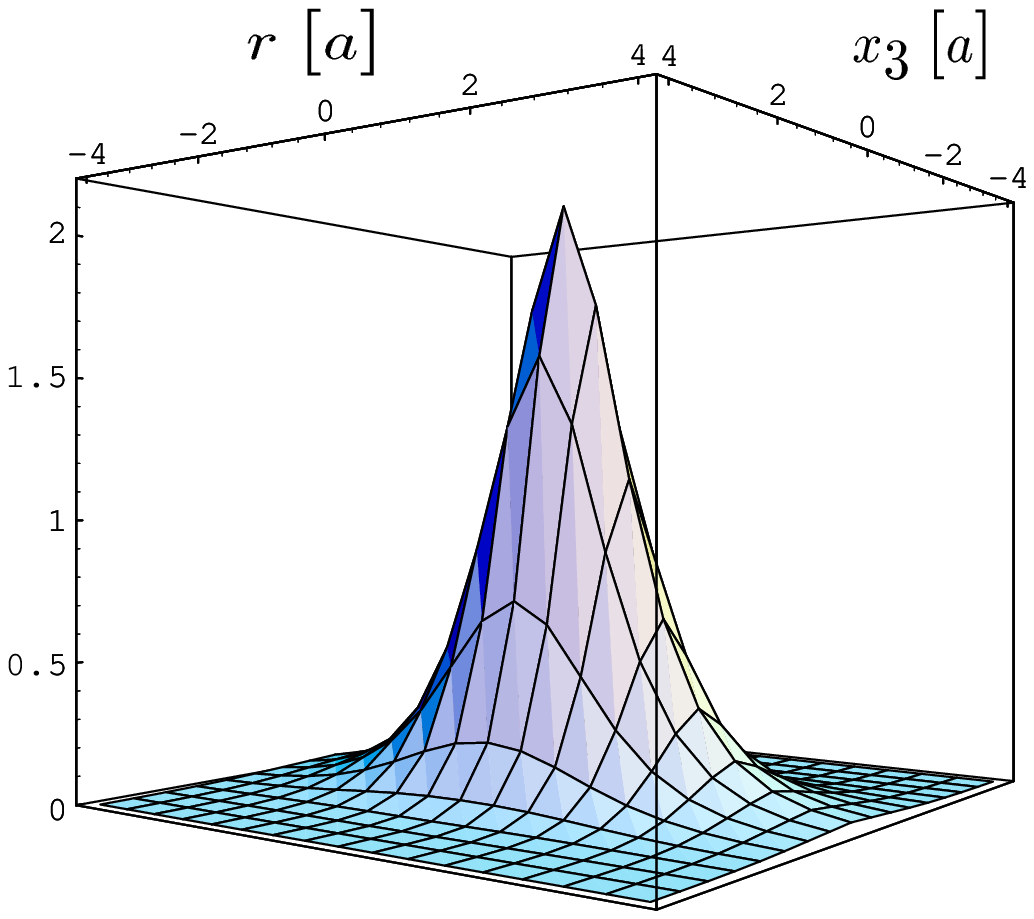}
\end{minipage}
\begin{minipage}{0.55\linewidth}
\epsfxsize0.7\linewidth
\epsfbox{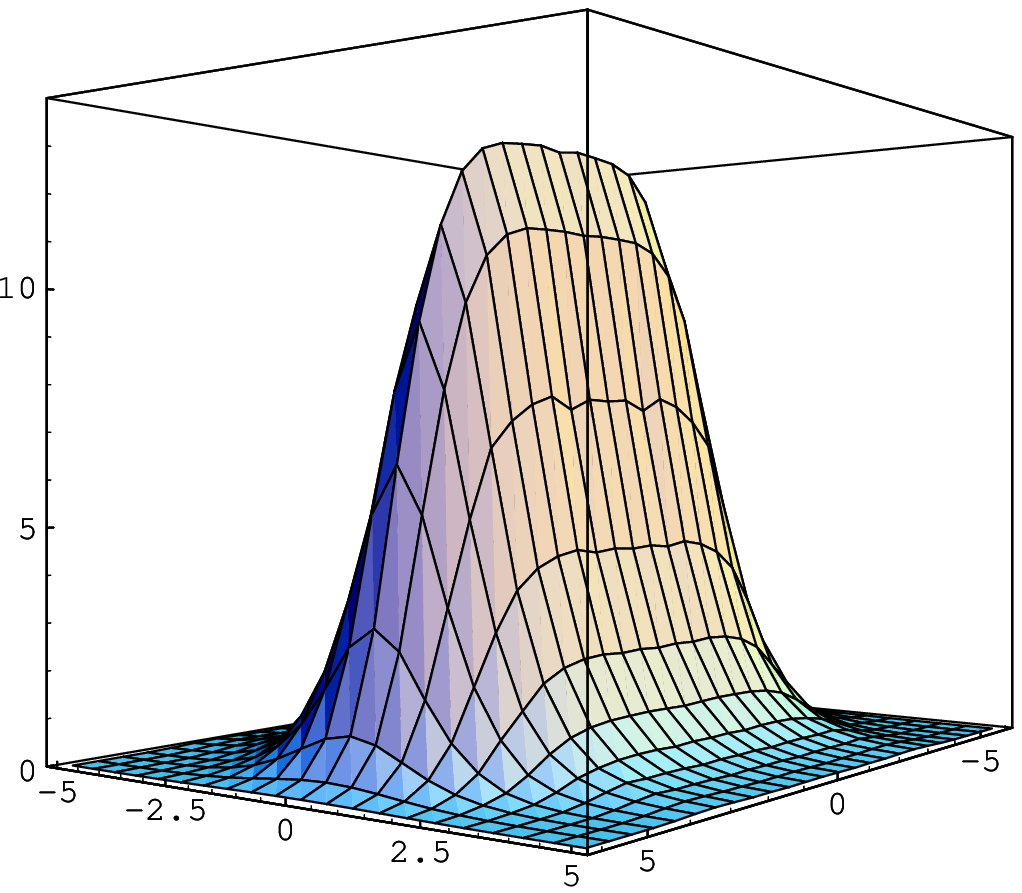}
\end{minipage}
\bigskip

\noindent {\bf Figs.6,7} Difference
of the (non-perturbative) squared electric field parallel to the quark
($x_3$) axis
\mbox{($-\lll E_z^2(x_3,r)\rrr_{\rm q \bar{q}-vac}$)} in $\rm
{\mbox{GeV}}/{\mbox{Fm}}^3$ for quark separation
$R_W$= 0.7 Fm ({\bf Fig.6}, left) and $R_W$= 3 Fm ({\bf Fig 7}, right);
$T_g\equiv a =0.35$ Fm (from \cite{stfrm}).

  \end{document}